\documentclass[aps,prl,superscriptaddress,twocolumn,longbibliography,letterpaper,accepted=2024-08-21]{quantumarticle} 

\pdfoutput=1 

\usepackage{graphicx}
\usepackage{amsmath,amsfonts,amssymb,amsthm}
\usepackage{color}
\usepackage{soul}
\usepackage{mathrsfs}
\usepackage{mathtools}
\usepackage{physics}

\usepackage[T1]{fontenc}
\usepackage[utf8]{inputenc}
\usepackage{babel}
\usepackage{booktabs}
\usepackage{tikz}
\usetikzlibrary{quantikz}
\usepackage[unicode=true,pdfusetitle,bookmarks=true,bookmarksnumbered=false,bookmarksopen=false,
 breaklinks=false,pdfborder={0 0 1},colorlinks=false]
 {hyperref}
\newcommand{\qvdots}{\raisebox{0.2em}{\ensuremath{\vdots}}}

\begin{document}

\title{Confinement and Kink Entanglement Asymmetry on a Quantum Ising Chain}

\author{B.~J.~J.~Khor}
\email{bk8wj@virginia.edu}
\affiliation{Department of Physics, University of Virginia, Charlottesville, VA, USA}
\affiliation{Fermi National Accelerator Laboratory, Batavia, IL 60510, USA}

\author{D.M.~K\"urk\c{c}\"uo\~glu}
\email{dogak@fnal.gov}
\affiliation{Fermi National Accelerator Laboratory, Batavia, IL 60510, USA}  
\affiliation{Superconducting Quantum Materials and Systems Center (SQMS),
Fermi National Accelerator Laboratory, Batavia, IL 60510, USA}

\author{T.~J.~Hobbs}
\email{tim@anl.gov}
\affiliation{High Energy Physics Division, Argonne National Laboratory, Argonne, IL 60439, USA}  

\author{G.~N.~Perdue}
\email{perdue@fnal.gov}
\affiliation{Fermi National Accelerator Laboratory, Batavia, IL 60510, USA}  

\author{I.~ Klich}
\email{klich@virginia.edu}
\affiliation{Department of Physics, University of Virginia, Charlottesville, VA, USA} 


\begin{abstract}
In this work, we explore the interplay of confinement, string breaking and entanglement asymmetry on a 1D quantum Ising chain. We consider the evolution of an initial domain wall and show that, surprisingly, while the introduction of confinement through a longitudinal field typically suppresses entanglement, it can also serve to increase it beyond a bound set for free particles. Our model can be tuned to conserve the number of domain walls, which gives an opportunity to explore entanglement asymmetry associated with link variables. We study two approaches to deal with the non-locality of the link variables, either directly or following a Kramers-Wannier transformation that maps bond variables (kinks) to site variables (spins). We develop a numerical procedure for computing the asymmetry using tensor network methods and use it to demonstrate the different types of entanglement and entanglement asymmetry.

\end{abstract}

\maketitle

\section{Introduction}

Strongly-coupled theories such as QCD can possess rich structure-forming properties relevant to many domains in
modern physics.
Despite advances in nonperturbative methods like lattice gauge theory and phenomenological modeling, a thorough understanding of QCD remains elusive due to the phenomenon of confinement~\cite{confinement1974wilson}. Recently, entanglement entropy has been suggested as providing a theoretical tool to investigate QCD systems, both in terms of bound states~\cite{Kharzeev:2017qzs,Lamm:2019uyc,Ehlers:2022oal} and with respect to scattering processes~\cite{Cervera-Lierta:2017tdt,Beane:2018oxh,Afik:2020onf,Liu:2022grf,Carena:2023vjc}.

In parallel, quantum spin chains have been proposed as analogous systems to investigate confinement analytically and numerically~\cite{mccoy1978two,delfino1996non,rutkevich1999decay,bhaseen2005aspects,rutkevich2008energy,Vovrosh:2020nwb, ranabhat2023dynamical}. In particular, the Ising spin chain has been a useful setup to study confinement in real time~\cite{kormos2017real}. Here, a two-fermion system is represented by domain walls, with binding effects introduced via a longitudinal field that gives an energy penalty linearly proportional to the length of the domain wall \cite{coldea2010quantum,zhang2020observation,zou20218}. This configuration simulates key aspects of the confinement of quark-antiquark pairs into mesons, or the binding of two-nucleon systems into the deuteron. This simple model, demonstrating confinement in non-equilibrium quantum quench dynamics, has led to an avalanche of related theoretical works~\cite{james2019nonthermal, robinson2019signatures, liu2019confined, verdel2020real, verdel2023dynamical, scopa2020entanglement, Castro-Alvaredo:2020prl, Milsted:2020jmf, mazza2019suppression, fragmentation2022bastianello, lerose2020quasilocalized, karpov2022spatiotemporal, pomponio2022bloch, lagnese2021false, lagnese2023detecting, knaute2024entanglement}, and the Ising chain confinement was also recently realized on IBM's quantum hardware~\cite{Vovrosh:2020nwb, Sopeno2021}.

The theory of many-body quantum entanglement can shed insights into confinement and symmetry breaking. Indeed, signatures of confinement show up in entanglement dynamics, where the entanglement entropy is greatly suppressed upon the introduction of the confining field \cite{kormos2017real, Vovrosh:2020nwb, scopa2020entanglement}. Symmetry breaking can also manifest in measures of entanglement entropy. When a wave function possesses a local symmetry, its entanglement entropy is a statistical combination of entropies associated to each local occupation number sector \cite{klich2008scaling, dasenbrook2015dynamical,barghathi2018renyi}. 
When a local symmetry is broken, however, additional contributions to entropy are generated and can be quantified using ``entanglement asymmetry''~\cite{ares2023entanglement, ares2023lack, ares2023non, capizzi2023entanglement}.
Various measures of symmetry-resolved entropy have been explored \cite{vittorio2022symmetry, rath2023entanglement, goldstein2018symmetry, murciano2022symmetry, murciano2020entanglement, bonsignori2019symmetry,vittorio2022symmetry, neven2021symmetry,murciano2023entanglement}. 

\section{Model and Entanglement Measures}

In this paper, we study the nature of meson dynamics through the lens of entanglement entropy and its asymmetry. In particular, we address the questions: is entropy always suppressed by a confining field? How can one address entanglement asymmetry for a quasi-local conservation law? To do so, we consider the Ising model with transverse and longitudinal fields and an additional three-body term that can be tuned to render the dynamics meson-number conserving for special points in parameter space~\cite{ge2023meson, iadecola2020quantum}:
\begin{align}
    H =\!-  J_0 \!\sum^{L-1}_{i=1}\! \sigma^z_i \sigma^z_{i+1} \!-\! g \!\sum^{L-1}_{i=2}\! \sigma^x_i \!-\! h \!\sum^{L}_{i=1} \!\sigma^z_i \!-\! J\! \sum_{i=1}^{L-2}\! \sigma_i^z \sigma_{i+1}^x \sigma_{i+2}^z\ .
\label{eq:TFIM}
\end{align}
This model exhibits confinement in the spreading of meson (kink/anti-kink pair) excitations for nonzero longitudinal fields, $h\!\neq\!0$~\cite{Vovrosh:2020nwb,Milsted:2020jmf,Vovrosh:2022bpj}. When we set $J\! =\! - g$, the model is dual to a fermionic chain coupled with a $\mathbb{Z}_2$ gauge theory \cite{borla2020confined}; it is kink-number preserving and exhibits quantum many body scars~\cite{ge2023meson, iadecola2020quantum,gustafson2023preparing}. We further note that the three-body term in the kink preserving regime naturally appears in applications such as the anti-blockade regime of the Rydberg simulators \cite{ostmann2019localization}, and that quantum many-body scar eigenstates of the kink preserving model have been prepared on IBM superconducting hardware \cite{gustafson2023preparing}.

The Hamiltonian \eqref{eq:TFIM} can be mapped to a fermionic model using a combination of Kramers-Wannier and Jordan-Wigner transformations (see Appendix E): 
\begin{eqnarray}
    \!H\! &=&  -  2 J_0\sum^{L}_{j=2} c_j^{\dagger} c_j -(g\!-\!J)\! \sum_{j=2}^{L-1} (c_{j}^{\dagger} c_{j+1} +h.c.) \nonumber \\ 
    &-&  (g\!+\!J)\! \sum^{L-1}_{j=2} (c_j^{\dagger} c_{j+1}^{\dagger} + h.c. ) - h \sum_{j=1}^L  \prod_{i=1}^j \left( 2 c_i^{\dagger} c_i\! -\! 1 \right). \label{eq:fermion_hamiltonian}
\end{eqnarray}
Here $c_j^{\dag}$ are fermion creation operators associated with creating a kink between site $j$ and $j-1$ in the original model. Eq. \eqref{eq:fermion_hamiltonian} shows that indeed kink-number preservation is exact for $J\! =\! - g$, allowing us to study regimes away from small $g,J$. On the other hand, when $g \neq -J$, only the kink-number parity is conserved.  Throughout the rest of the paper we set $J_0 = 1, \hbar=1$, thus  time scales throughout this paper are set in the unit of $J_0^{-1}$.

We consider the time evolution of an initial domain wall product state of length $n$, i.e., of the form 
\begin{eqnarray}
    | j, n \rangle \equiv | \dots \uparrow \downarrow_j \dots \downarrow_{j+n-1} \uparrow \dots  \rangle\ .
\label{eq:domain wall basis}
\end{eqnarray}
Entanglement can be quantified via R\'enyi entropies,
\begin{equation}
S_n (\rho_A) = \text{log}_2(\text{Tr}(\rho_A^n))/(1-n)\ ,
\label{eq:renyi}
\end{equation}
where $\rho_{A}=\text{Tr}_B \rho $ is a reduced density matrix associated with a subset of sites $A$ when its complement, $B$, has been integrated out. In particular, the second-order R\'enyi entropy is $S_{2}(\rho_A) = -\log_2 \mathrm{Tr}(\rho^2_A)$.

We evaluate $ S_{2}(\rho_A)$ in numerical simulations below. Within time-evolving block decimation (TEBD)~\cite{white2004real, efficient2004vidal, matrix2004verstraete}, this generalizes to other entropy orders. Another quantity of interest is the kink density, $\Delta^{zz}_{i,i+1} \equiv {1 \over 2}\langle   (1-\sigma^z_i \sigma^z_{i+1})  \rangle$, with $\Delta^{zz}_{i,i+1} = 1\, (0)$ for a spin flip (alignment).  

\section{Kink-preserving Dynamics}

As mentioned, with the choice $J\!=\!-g$, the Hamiltonian in Eq.~\eqref{eq:TFIM} preserves kink number. To study the evolution of a kink, we project $H$ onto the two-kink subspace in the kink basis, Eq.~\eqref{eq:domain wall basis}:
\begin{eqnarray}
 &   H_{2} = \!\!\!\sum\limits_{\substack{\scriptscriptstyle 0 \leq j < L-1 \\ \scriptscriptstyle 0 < n < L - j - 1}} \!\!\!  2 h n |j, n \rangle \langle j, n | \!-\! \nonumber  (g - J) \Big( | j-1, n+1 \rangle +  \\ & 
 | j+1, n-1 \rangle 
    + | j, n-1 \rangle + | j, n+1 \rangle \Big) \langle j, n |\ .
\label{eq:two-kinks}
\end{eqnarray}
This projection is exact when $H$ conserves kink number. Eq. \eqref{eq:two-kinks} highlights an advantage of adding the three body term in controlling kink production compared to working without the three body interaction $\sigma_i^z \sigma_{i+1}^x \sigma_{i+2}^z$ modification of the transverse field Ising, while assuming $g$ is small. Indeed in the later case, kinks appear perturbatively on the background of a classical Ising chain and propagate slowly, since their kinetics is governed by $g$. Including the three body term, Eq. \eqref{eq:two-kinks} shows that the dynamics of the two kink bound states (with $g=-J$) is characterized by the kinetic energy scale $g-J$, which can be tuned to be arbitrarily large, allowing us to study regimes of fast dynamics while preserving kink number. 

We note that the exactness of $H_2$ offers an opportunity to benchmark tensor network methods against the corresponding computation by exact evolution in the two-kink subspace. In particular, we compare the evolution of the local kink density, as well as the 2$^{\text{nd}}$ R\'enyi entropy as computed by exact diagonalization within the two-kink subspace with corresponding computations within TEBD evolution. Interestingly, we are not aware of any previous calculation of R\'enyi entropy directly in the two-kink subspace Hamiltonian in Eq.~\eqref{eq:two-kinks}, and we outline the computational approach in Appendix B. We also verify the validity of the two-kink subspace dynamics by comparing the time evolution for various observables and the R\'enyi entropy with identical quantities computed from small-size exact diagonalization in Appendix A. 

\begin{figure}[h]
\centering
\includegraphics[width=0.5\textwidth]{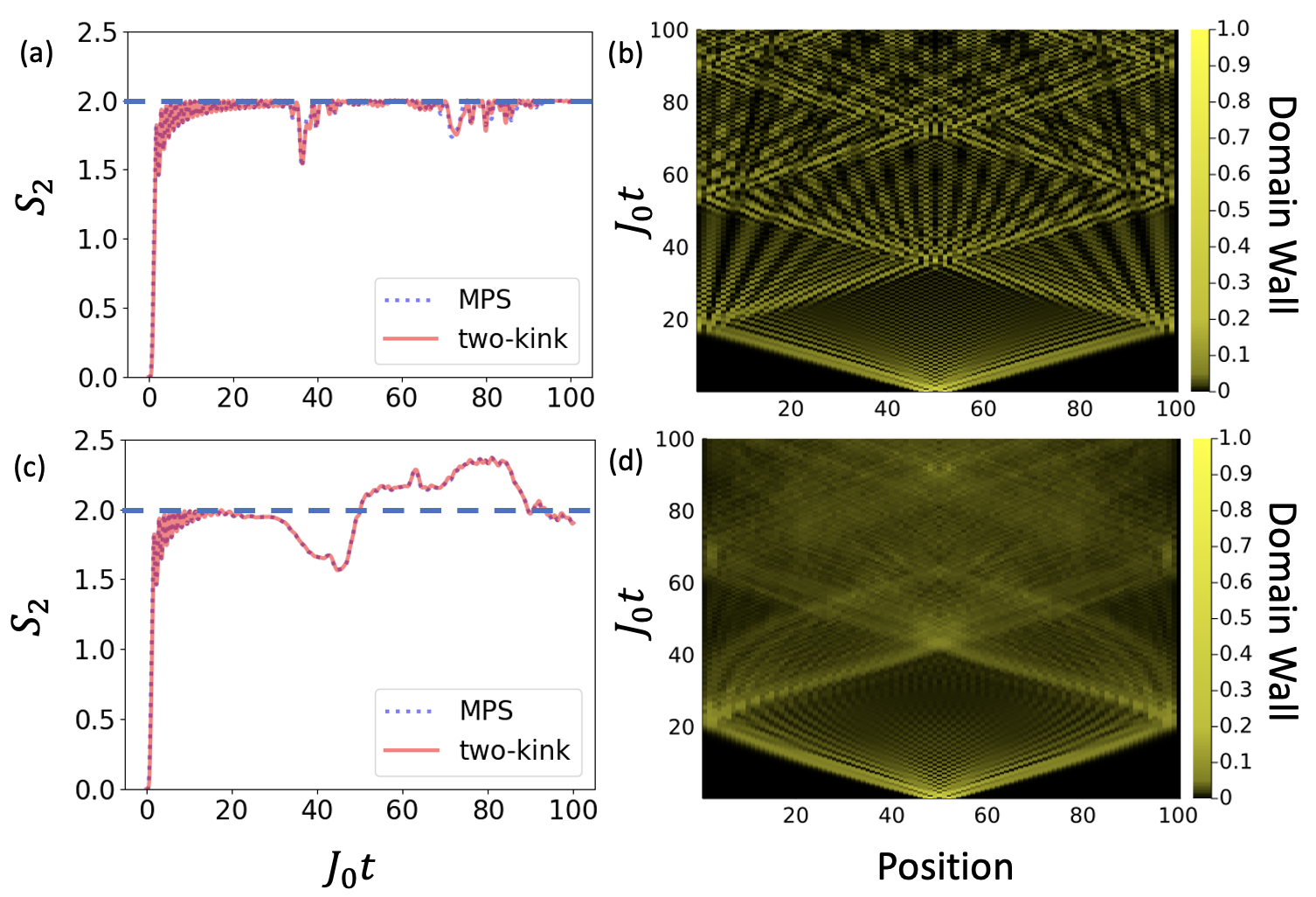}
  \caption{Evolution of half chain R\'enyi entropy for (a) $g =-J= 0.7, h = 0$ and (c) $g =-J= 0.7, h = 0.05$. The corresponding domain wall profile following a quench are (b) $g =-J= 0.7, h = 0$ and (d) $g = -J = 0.7, h = 0.05$ respectively. The initial state $| \uparrow \dots \uparrow \downarrow \downarrow \downarrow \downarrow \uparrow \dots \uparrow \rangle$ is $L = 100$ spin chain with $n=4$ initial domain size in the middle of the spin chain.}
\label{fig:kink-preserving}
\end{figure}

\begin{figure}[h]
\centering
\includegraphics[width=0.5\textwidth]{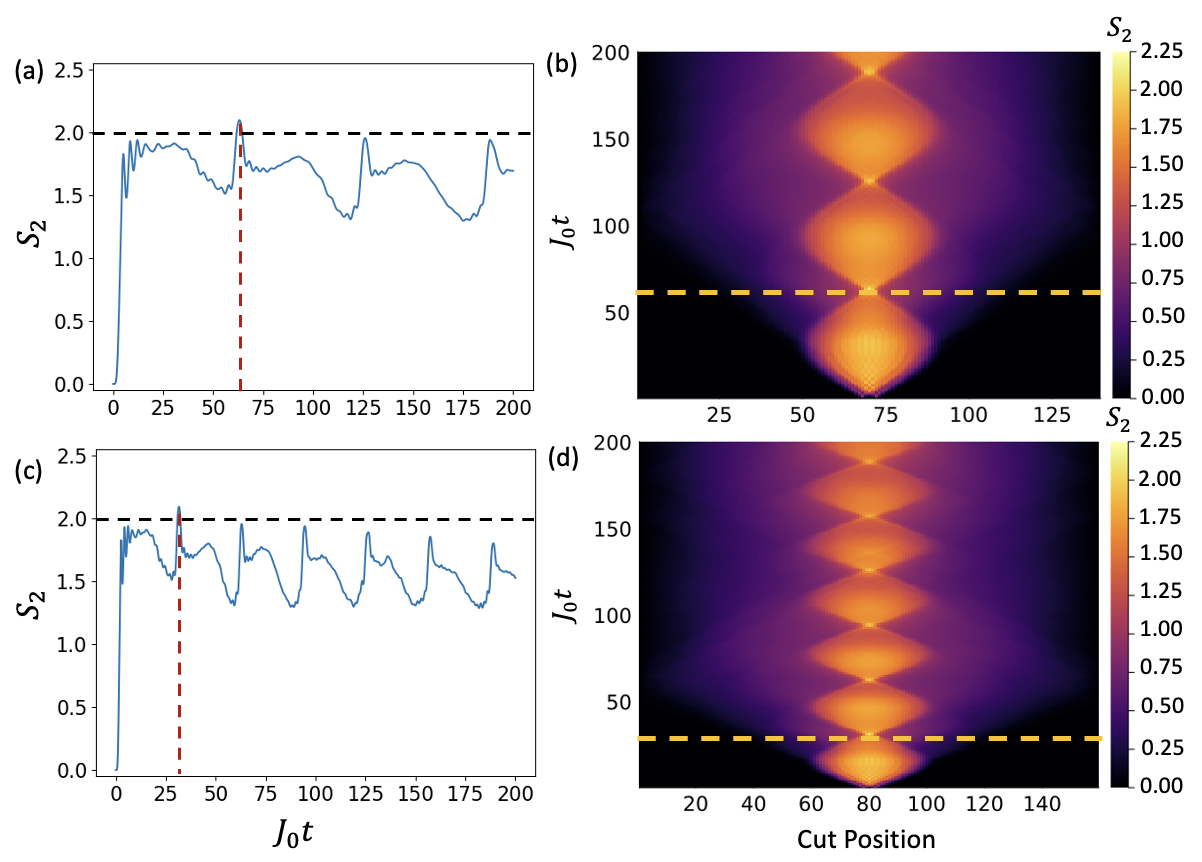}
    \caption{Half chain R\'enyi entropy for (a) $g =-J= 0.25, h = 0.05, L=140$ and (c) $g = -J = 0.5, h = 0.1, L =160$. The right figures are bipartite R\'enyi entropy at all possible cuts along the chain for (b) $g =-J= 0.25, h = 0.05, L=140$ and (d) $g = -J = 0.5, h = 0.1, L =160$, for which the left figures are only a slice along the middle of the spin chain. The initial states $| \uparrow \dots \uparrow \downarrow \downarrow \downarrow \downarrow \uparrow \dots \uparrow \rangle$ has $n=4$ initial domain size in the middle of the spin chain of size of either $L = 140$ or $L = 160$. See Appendix C for accompanying kink density plot.}
\label{fig:renyi_map}
\end{figure}

\section{R\'enyi Entropy saturation,  integrability and confinement}

In Fig.~\ref{fig:kink-preserving}, we consider the evolution of an initially small domain, focusing on the half-chain entropy, $S_2 (t)$, and on the kink density.
Note the remarkable agreement between the exact diagonalization and TEBD computations, validating the TEBD approach, at least for small kink numbers.

A striking feature of the R\'enyi entropy dynamics in Fig.~\ref{fig:kink-preserving} (a) is that when $h=0$ and $g=-J$, the second-order R\'enyi entropy saturates at $S_2 \leq 2$ when we set our initial state to be a two-kink state. This is surprising at first since a generic two-kink wave function can have a much larger entropy, up to $\log_2(L/2+2)$. In Appendix F, we prove that if $h=0$, and given an initial two-kink state, we have $S_2 \leq 2$ when kink number is conserved. 

This bound can be roughly understood as follows.  {An initial domain wall state \eqref{eq:domain wall basis} corresponds, in the fermion formulation \eqref{eq:fermion_hamiltonian}, to two initially localized fermions 
\begin{eqnarray}
  c_j^{\dagger }c_{j+n}^{\dagger
   }|\text{vac}\rangle. 
\end{eqnarray} 
When we take $h = 0, J = -g$, the second line in \eqref{eq:fermion_hamiltonian}, vanishes leaving us with a free fermion Hamiltonian. The subsequent evolution is of a pair of non-interacting, uncorrelated,  fermions.  Intuitively, each fermion can contribute at most $S_2 = \mathrm{log}_2(2) = 1$ due to its de-localization in the system. The uncorrelated nature of the fermions then implies $S_2 \leq 2$. A rigorous proof is given in Appendix F.

\subsection{The effect of confinement and integrability breaking}

What happens when $h >0$? Naively, one may expect that entropy generation will be decreased, due to the reduction of the spread of the particles. 

However, considering Fig.~\ref{fig:kink-preserving} (lower), we encounter a surprise: with a small $h=0.05$, we find a clear violation of the free-particle ($h\!=\!0$) bound. Indeed, the collisions and interactions  with $h\neq 0$ do not correspond to non-interacting particles and are not bound by the above argument, { as can be seen in the fermionic model \eqref{eq:fermion_hamiltonian} where the confining term $h$ gives rise to a highly non-local interaction term, thereby allowing quantum correlations and entanglement to develop}. We explore this more closely in Fig.~\ref{fig:renyi_map}. The evolution exhibits the following features: when the string reaches its maximal extent, the entropy starts decreasing, due to a suppression of the wave function spread. However, close to the minimal string size, where collision is possible, entropy shows a rapid increase. When this collision-related increase is sufficient to overcome suppressed entropy, the $S_2\!=\!2$ bound is violated. 

Note that in Fig.~\ref{fig:kink-preserving}, the bound violation is enhanced {when the size of the chain is smaller than the maximal extent of the wave function spread}. In Fig.~\ref{fig:renyi_map} and the accompanying kink density plot in Appendix C, however, we find that early time collisions can still violate the bound before these boundary effects become substantial, showing that the bound violation may persist also in the thermodynamic limit, $L \rightarrow \infty$. The variations in the R\'enyi entropy of the system at all cut positions are plotted in the right panels of Fig.~\ref{fig:renyi_map}, showing that the oscillatory behavior of the entropy is reproduced along the internal structure of the meson.

\begin{figure}[h]
\centering
\includegraphics[width=0.4\textwidth]{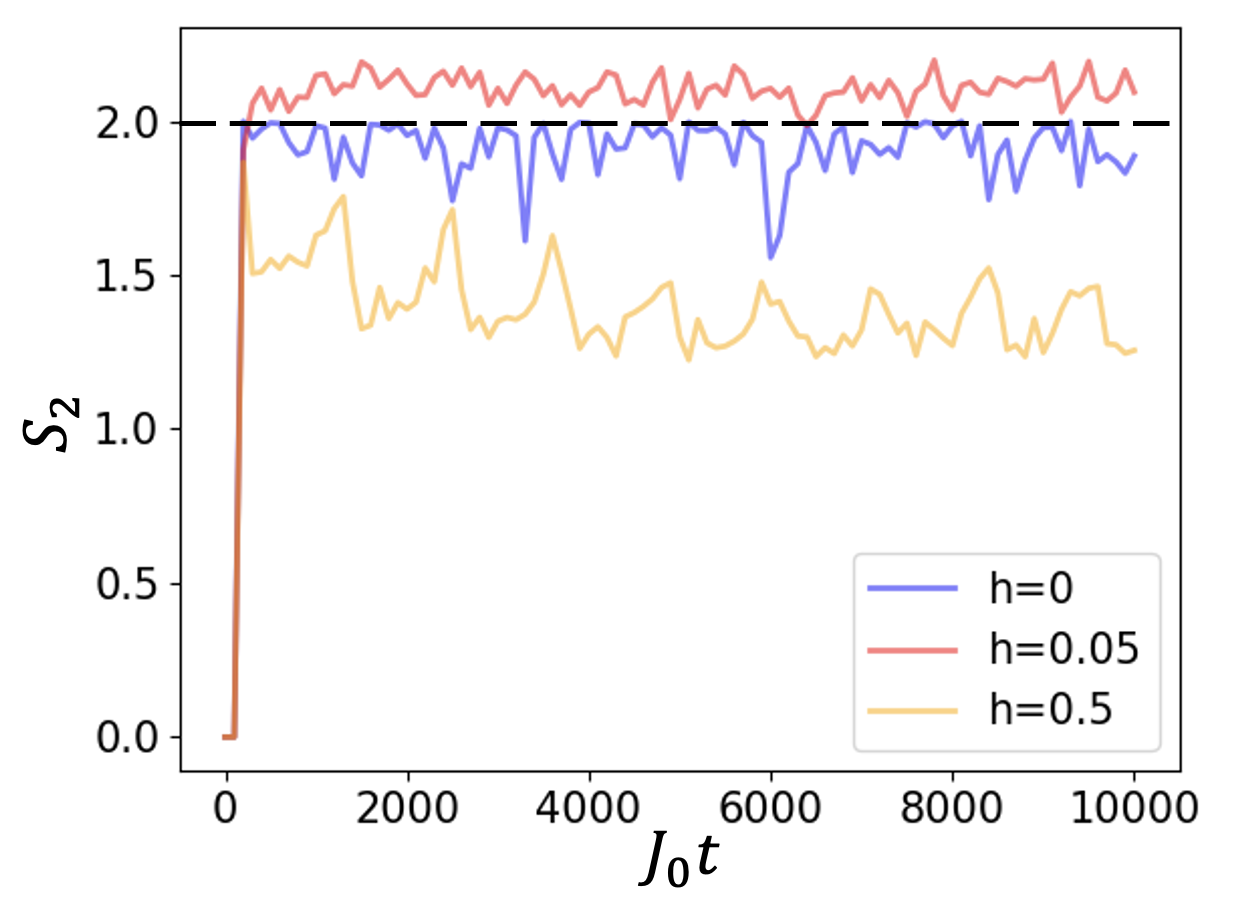}
    \caption{The long time dynamics of the bipartite R\'enyi entropy $S_2(t)$ from the exact diagonalization of the two-kink Hamiltonian $H_2$ in Eqn \ref{eq:two-kinks}. Here, the kink kinetic energy parameter is set to $g = -J=0.7$, the chain is of length $L = 100$, and we notice that for weak confining field $h$, the entropy exceeds the $S_2 = 2$ limit while strong confining field $h$ eventually suppresses the entropy growth.}
\label{fig:late_time}
\end{figure}

In Fig.~\ref{fig:late_time} we take advantage of the fact that, in exact diagonalization, no Trotterization error is introduced and we can simulate our time evolution dynamics to arbitrarily late times in the exact kink-preserving simulation. Consistent with Fig.~\ref{fig:kink-preserving}, when we have $h = 0$, the R\'enyi entropy is bounded by $S_2\! \leq\! 2$. Upon introducing a small confining field, we see that the R\'enyi entropy can exceed this upper bound. On the other hand, once the confining field becomes significantly stronger ($h = 0.5$), the R\'enyi entropy is suppressed compared to the unconfined case, and exhibits oscillatory behavior.

\section{Entanglement Asymmetry}

We now turn to our second question. What is the effect of symmetry breaking on entanglement entropy when the symmetry operator is not exactly local? Given a local symmetry operator of the form $Q=Q_A\otimes I+I\otimes Q_B$ such as charge or magnetization, a useful quantity utilized to study symmetry breaking and its relation to entanglement entropy is the entanglement asymmetry. 

The entanglement asymmetry is obtained as follows. First, we project $\rho_A$ onto the blocks associated with different symmetry sectors of $Q_A$:
\begin{equation}
    \rho_{A,Q} = \sum_q \Pi_{q} \rho_A \Pi_{q}\   ,
\end{equation}
where $\Pi_{q}$ are projectors onto a subspace with a given eigenvalue $q$ of $Q_A$. When $Q_A$ has integer eigenvalues, the projected density matrix can be written as
\begin{equation}
    \rho_{A,Q} = \int_{-\pi}^{\pi} \frac{d \lambda}{2 \pi}  e^{- i \lambda Q_A} \rho_A e^{i \lambda Q_A}\ . \label{eq:projected matrix}
\end{equation}

Then, the projected density matrix can be used to construct the entanglement asymmetry \cite{ares2023entanglement}, given by
\begin{equation}
    \Delta S_n(\rho_A) = S_n (\rho_{A, Q}) - S_n (\rho_A)\ ,
    \label{eq:entanglement asymmetry} 
\end{equation}
where $S_n (\rho_{A, Q}) = \log_2 \mathrm{Tr} (\rho_{A,Q}^n)/(1-n)$ is the symmetry-resolved R\'enyi entropy. The entanglement asymmetry vanishes, $\Delta S_n = 0$, if computed for a state that commutes with $Q$. A more detailed introduction to entanglement asymmetry is given in, e.g.  \cite{ares2023entanglement}.

\subsection{Kink and Kramers-Wannier Entanglement Asymmetries}

In contrast to the discussion above and also examples presented in existing literature, { we ask the question as to how we can treat entanglement asymmetry when the symmetry operator is not fully local, i.e., the operator associated with the symmetry is not a sum of single site operators.} 

In our current context, the conservation of kink number is associated with a quasi-local charge (a two-body operator in our case), $N_k$, since kinks live on the dual lattice, with the number of kinks given by
\begin{eqnarray}
    N_k=\sum^{L-1}_{i=1} \Delta^{zz}_{i,i+1}=\frac{1}{2}\left(L-1-\sum^{L-1}_{i=1} \langle \sigma^z_i \sigma^z_{i+1} \rangle\ \right) .\label{eq:kink_num_op}
\end{eqnarray}
We break the spin chain into complementary subsystems, $A$ and $B$, where $A$ contains spins $1,..,L_A$; then,
\begin{eqnarray}
    N_k=N_{k,A}\otimes I+I\otimes N_{k,B}+\Delta^{zz}_{L_A,L_A+1}\ , 
    \label{eq:kink_bipartition}
\end{eqnarray}
where $N_{k,A},N_{k,B}$ count the number of kinks within $A$ and its complement $B$ respectively. The last term measures the presence of a kink at the interface between $A$ and $B$. 

The presence of a kink at the interface between $A$ and $B$ is impossible to determine from within subsystem $A$ alone. However, we may compute a coarse-grained entanglement asymmetry by projecting onto blocks with fixed kink number inside $A$. We define the projected density matrix, $\rho_{A, N_{k,A}}$, as
\begin{eqnarray}
 \rho_{A, N_{k,A}}=\int_{-\pi}^\pi \frac{{\rm d}\lambda }{2\pi}e^{-i\lambda N_{k,A}}\rho_A e^{i\lambda  N_{k,A}}\ .
 \label{eq:kink_projected_rrho}
\end{eqnarray}
This projection is depicted schematically in Fig.~\ref{fig:reduced density matrix kink blocks.}; from this, we define an asymmetry, $\Delta S_2^{kink}\equiv \Delta S_2( \rho_{A, N_{k,A}})$, as in Eq.~\eqref{eq:entanglement asymmetry}.
\begin{figure}[h]
\centering
\includegraphics[width=0.4\textwidth]{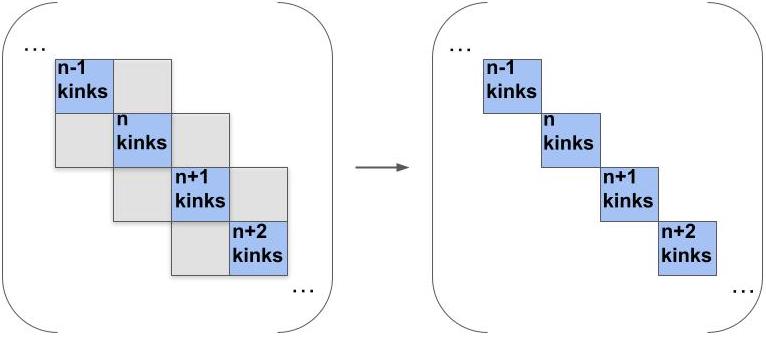}
    \caption{Left: The reduced density matrix $\rho_A$ for an initial wave function with a fixed number of kinks may have terms coupling $n$ kinks in the bulk of $A$ with $n+1$ or $n-1$ kinks. Right: the projected matrix $ \rho_{A, N_{k,A}}$.}
\label{fig:reduced density matrix kink blocks.}
\end{figure}

We note that asymmetries computed using $\rho_{A, N_{k,A}}$ may be nonzero even if the overall wave function has a fixed kink number. { This is in contrast with the entanglement asymmetry $\Delta S_n$ associated with strictly local (one-body) symmetry operator, where $\Delta S_n$ necessarily vanishes when the wave function has fixed symmetry eigenvalue. In the case of quasi-local (two-body) kink operator, owing to the presence of the kink at the interface, $\Delta S_n$ can be nonzero even when the wave function has fixed kink number.}

To see this, consider the example of a very simple wave function:
\begin{eqnarray}
   {1\over \sqrt{2}} (| \uparrow \uparrow \downarrow \downarrow \rangle
+| \uparrow \downarrow \downarrow \downarrow \rangle)\ ,
\label{eq:example kink conserve}
\end{eqnarray}
where the first two spin sites define subsystem $A$, giving:
\begin{eqnarray}
    \rho_A=\frac{1}{2}(|\uparrow \uparrow \rangle
   +|\uparrow \downarrow \rangle
   )(\langle \uparrow \uparrow
   |+\langle \uparrow \downarrow |)\ .
\end{eqnarray}
Note that $\rho_A$ describes a pure state with no entropy, however it is {\it not} in block form from the point of view of internal kinks in subsystem $A$. On the other hand
\begin{eqnarray}
    \rho_{A,N_{k,A}}={1\over 2}\big(|\uparrow \uparrow \rangle
   \langle \uparrow \uparrow
   |+|\uparrow \downarrow \rangle
   \langle \uparrow \downarrow |\big)
\end{eqnarray}
describes a mixed state with entropy $\log_2 2=1$. 

Thus our generalized entanglement asymmetry will only vanish for wave functions that have a fixed number of kinks and no kink at the interface between $A$ and $B$. The presence of kink at the interface, even with fixed kink number, leads to nonzero asymmetry. Nevertheless, the contribution to the asymmetry, $\Delta S_2^{\textrm{kink}}$, from the boundary kink is small: it will be responsible for at most an $O(1)$ contribution to the entropy, and is thus suitable for probing entropy scaling in large systems. 

{ How can we take into account the effect of a possible kink lying at the interface of the bipartition when one calculates quantities such as the entropy?} Here, we propose an asymmetry measure that vanishes for eigenstates having total kink number $N_k$. To do so, we use an open boundary Kramers-Wannier (KW) transformation $U_{KW}$, which maps $|s_1,..s_N\rangle \rightarrow |t_1,..t_N\rangle$ where $t_1=s_1$ and $t_i=s_{i-1}s_i$ for $i>1$. { The Kramers-Wannier transformation maps the link variable (in our case, the kink) to a local site variable, i.e., $\sigma_i^z \sigma_{i+1}^z \rightarrow \sigma_{i+1}^z$, and naturally deals with the kink lying in the interface of the bipartition.} In particular, $U_{KW}$ maps kink number into magnetization which is completely local, { thus in the Kramers-Wannier (KW) basis, the magnetization entanglement asymmetry should vanish for a wave function with a fixed kink number.}

Let us consider the relationship between entropies in the original basis vs the KW basis. We take as the set of sites $A=1,..,L_A$ to be fixed. Given a quantum state on the full system we can define $\rho^{KW}=U_{KW} \rho U_{KW}^{\dagger}$, and $\rho^{KW}_{A}=\text{Tr}_B \rho_{KW}$. We show in appendix E that $U_{KW}$ only contains a single two-qubit gate operating between subsystems $A,B$, and hence $S(\rho_A)$ and $S(\rho^{KW}_{A})$ differ by at most the entropy that can be generated by such a gate, i.e. at most $2\log_2 2$. Therefore we can use $S(\rho^{KW}_{A})$ as an alternative measure for questions of entanglement scaling, i.e., whether entropy is bounded when $A$ is large. 

Next, let us discuss the transformation of the kink entanglement asymmetry in the original basis into the magnetization entanglement asymmetry in the Kramers-Wannier basis. Note that under $U_{KW}$, we have $\prod_{i=1}^{l-1} e^{\pm i \lambda \sigma_i^z \sigma_{i+1}^z} \rightarrow \prod_{i=2}^{l} e^{\pm i \lambda \sigma_i^z}$ (see Appendix E). Thus, following Eqs. \eqref{eq:projected matrix},\eqref{eq:kink_projected_rrho}, we define the KW projected density matrix as:
\begin{eqnarray}
    \rho^{KW}_{N_k,A}= \int_{-\pi}^\pi \frac{{\rm d}\lambda}{2\pi}e^{-i\frac{\lambda}{2} \sum_{l=2}^{L_A}\sigma^{z}_l}\rho^{KW}_{A} e^{i\frac{\lambda}{2} \sum_{l=2}^{L_A}\sigma^{z}_l}, \label{eq:kink_projected_KWrrho}
\end{eqnarray}
and the KW kink asymmetry as $\Delta S_2^{KW}\equiv S_2( \rho^{KW}_{N_k,A})-S_2( \rho^{KW}_{A})$. Note that, $\Delta S_2^{KW}$ obeys the desired property that $\Delta S_2^{KW}=0$ if the system is in a state with a fixed number of kinks. { Even in cases when kink number is not conserved in the original basis, we in general expect $\Delta S_2^{KW} \leq \Delta S_2^{kink}$, as the kink entanglement asymmetry in the original basis can miscount a kink lying at the interface [see Eq.~\eqref{eq:kink_bipartition}].}

\subsection{MPS implementation} 

To proceed beyond the kink conserving dynamics we have developed a Matrix Product State (MPS) procedure to compute $S_2$. Given a local charge of interest (magnetization or $N_{k,A}$) we compute the asymmetry by expressing $S_2 (\rho_{A, Q})$ as
\begin{equation}
    S_2 (\rho_{A, Q}) = \int_{0}^{2 \pi} \frac{d \lambda}{ \pi}\big(1-\frac{\lambda}{2 \pi}\big) \text{Tr}_A [e^{i \lambda Q_A} \rho_A e^{-i \lambda Q_A} \rho_A ] .
    \label{eq:second_EA}
\end{equation}
We carry out the trace in Eq.~\eqref{eq:second_EA} by representing $\rho$ as a matrix product operator and doing the necessary contractions. A discussion of the computational complexity of the procedure is provided in Appendix D. 

\begin{figure}[h]
\centering
\includegraphics[width=0.5\textwidth]{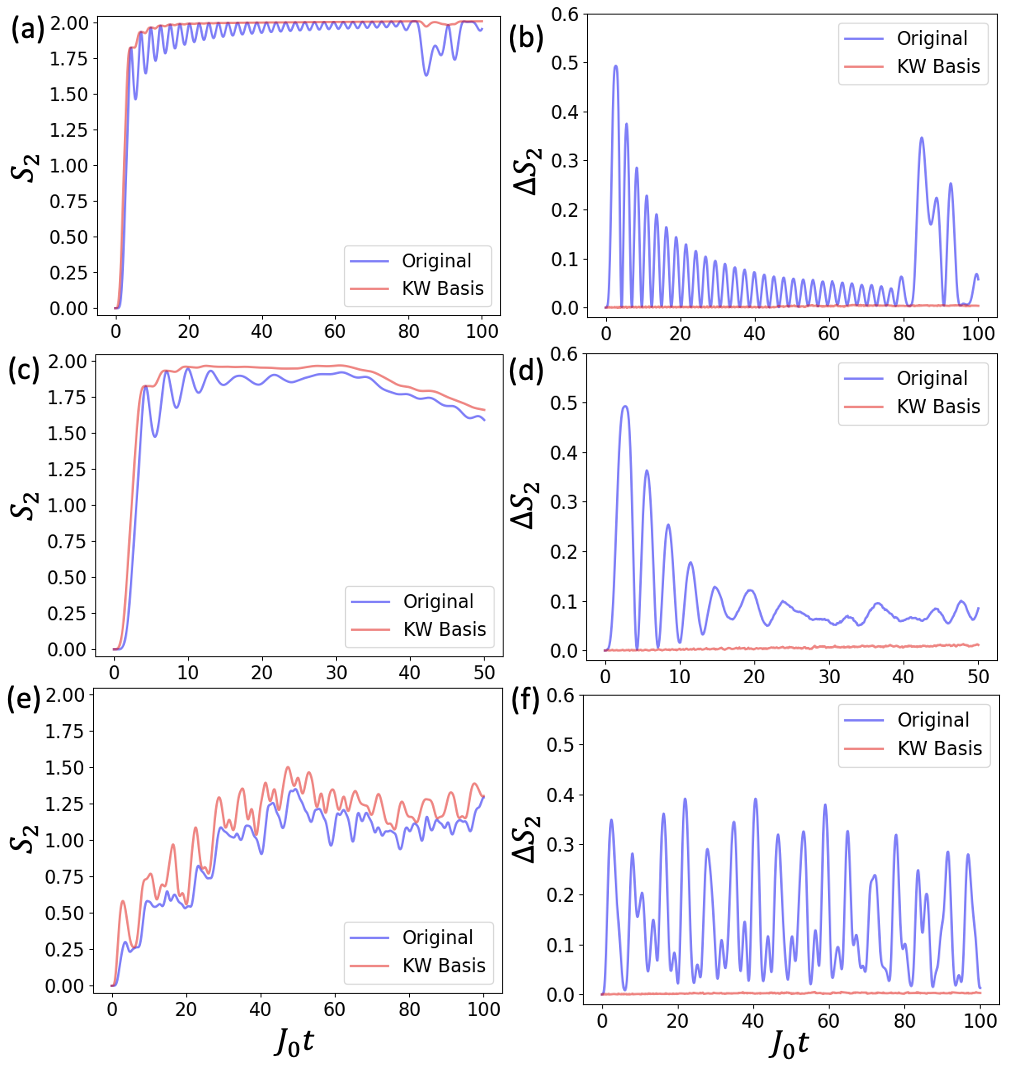}
    \caption{Entropy $S_2 (t)$ and asymmetry $\Delta S_2 (t)$ for the original basis vs Kramers-Wannier basis. Here, (a) $S_2(t) \text{ for } h = 0$, (b) $\Delta S_2(t) \text{ for } h = 0$ (c) $S_2(t) \text{ for } h = 0.05$, (d) $\Delta S_2(t) \text{ for } h = 0.05$, (e) $S_2(t) \text{ for } h = 0.5$, and (f) $\Delta S_2(t) \text{ for } h = 0.5$. Here, the transverse field and three-body strength are set at $g = 0.3 = -J$ for all figures. The initial states $| \uparrow \dots \uparrow \downarrow \downarrow \downarrow \downarrow \uparrow \dots \uparrow \rangle$ for $L = 100$ spin chain has $n=4$ initial domain size in the middle of the spin chain and the evolution is kink conserving. Note that the shorter time scale of (c) and (d) is due to the higher computational complexity to simulate the R\'enyi asymmetry for the case $h = 0.05$.}
\label{fig:Asymmetry_in_preserving_cases.}
\end{figure}

\begin{figure}[h]
\centering
\includegraphics[width=0.5\textwidth]{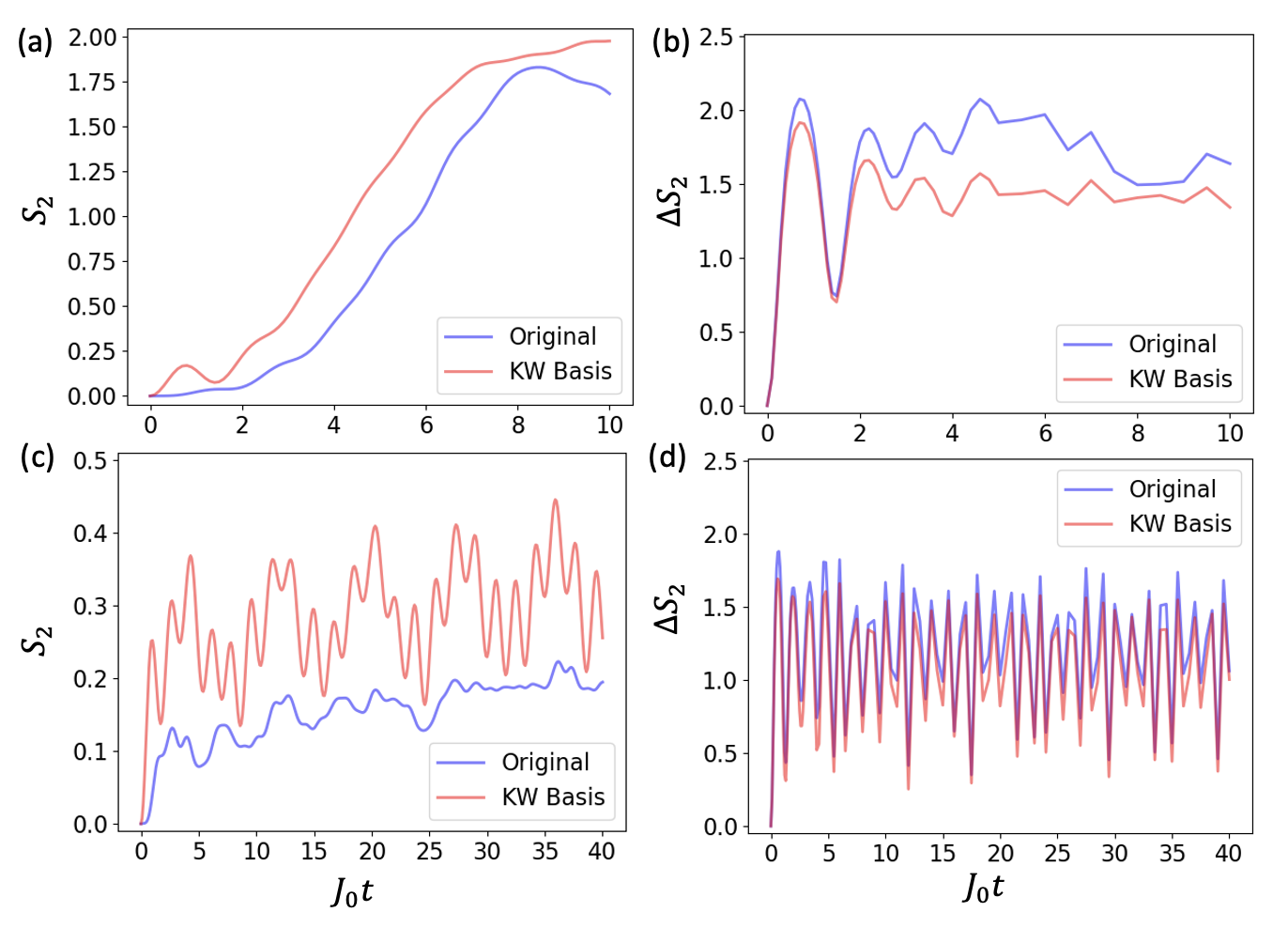}
    \caption{Entropy $S_2(t)$ and asymmetry $\Delta S_2 (t)$ in the original vs. the Krammers-Wannier basis.Here, (a) $S_2(t) \text{ for } h = 0$, (b) $\Delta S_2(t) \text{ for } h = 0$ (c) $S_2(t) \text{ for } h = 0.4$ and (d) $\Delta S_2(t) \text{ for } h = 0.4$. Here, the transverse field and three-body strength are set at $g = 0.4$ and $J = 0.1$ respectively for all figures. The initial states $| \uparrow \dots \uparrow \downarrow \downarrow \downarrow \downarrow \uparrow \dots \uparrow \rangle$ for $L = 60$ spin chain has $n=4$ initial domain size in the middle of the spin chain and the evolution is not kink conserving. Note that the short time scale of the asymmetry evolution owes to the limit of the computational ability to simulate the R\'enyi asymmetry at later time scales, especially for the $h=0$ case.}
\label{fig:Asymmetry_in_violation_cases.}
\end{figure}

\section{Numerical results}

We present the numerical results for both the kink-preserving cases and the string breaking cases in Figs.~\ref{fig:Asymmetry_in_preserving_cases.} and \ref{fig:Asymmetry_in_violation_cases.} respectively, in both the original basis and the KW basis, and for R\'enyi entropy $S_2$ and R\'enyi asymmetry $\Delta S_2$. We begin by commenting on the kink-preserving case $g = -J$ in Fig.~\ref{fig:Asymmetry_in_preserving_cases.}. Note that the kink R\'enyi asymmetry $\Delta S_2^{kink}$ does not vanish even when the kink number is conserved, consistent with our example illustrated earlier Eq.~\eqref{eq:example kink conserve}. On the other hand, we verify that in the KW basis the asymmetry $\Delta S_2^{KW}$ vanishes, as expected \footnote{Due to the approximate nature of the TEBD time evolution on the MPS, we note that there is error of order $10^{-3}$ on the kink number conservation, and the magnetization asymmetry error is fluctuating in order of $10^{-3}$.}.

In Fig.~\ref{fig:Asymmetry_in_preserving_cases.}, we study three different cases with different confining field strength while keeping the kinetic energy $g=-J$ constant: (1) the free fermion case $h = 0$, (2) weak confinement, $h = 0.05$, and (3) strong confinement $h = 0.5$. We note that $S_2( \rho^{KW}_{A})$ and $S_2( \rho_{A})$ are quite close to each other. { Moreover,  we observe that  $S_2 (\rho^{KW}_{A}) > S_2 (\rho_{A})$ consistently. We can explain this feature as follows. For weakly correlated states, with low entropy, the wave-function in the vicinity of the cut region is close to a product state in the original basis. As mentioned, the transformation $U_{KW}$ (see appendix E) contains an entangling 2-qubit gate (a CNOT) acting between the sides of the system, which will generically increase the entropy when acting on a weakly correlated state.
Thus, we expect that the R\'enyi entropy $S_2$ will be enhanced after the Kramers-Wannier transformation compared to the R\'enyi entropy $S_2$ in the original basis.} 

Note the pronounced oscillation of the entropy in the strong confining case. Due to the strong confinement, the kink is oscillating near the entropy cut, which is reflected in large oscillations of the asymmetry in the original basis, while, consistent with our construction, showing that the KW asymmetry vanishes $\Delta S_2^{KW} = 0$ to a good approximation. 

Finally, note that the dip in R\'enyi entropy in the original basis around $t \approx 85$ for $h = 0$ in Fig.~\ref{fig:Asymmetry_in_preserving_cases.} is associated with collision of kinks, and is accompanied with a spike in kink asymmetry in the original basis. In general, we observe that dips in $S_2$ corresponds to spikes in kink asymmetry $\Delta S_2^{kink}$. A kink density heatmap is provided in appendix C (see Fig \ref{fig:collision 1}) for reference. 

Next we consider kink number violating dynamics in Fig.~\ref{fig:Asymmetry_in_violation_cases.}. A common feature shared by the R\'enyi entropy in the kink-preserving Fig.~\ref{fig:Asymmetry_in_preserving_cases.} and string breaking Fig.~\ref{fig:Asymmetry_in_violation_cases.} is that the R\'enyi entropy in the original basis is in general lower than R\'enyi entropy in the KW basis.  

On the other hand, the kink entanglement asymmetry $\Delta S_2^{kink}$ is greater than the KW kink asymmetry $\Delta S_2^{KW}$, though the difference is more pronounced in the kink-preserving case. This behavior may be attributed to the entropy associated with the possibility of a kink at the center of the chain, exactly at the boundary between the left and right regions, as outlined in our discussion in earlier section justifying $\Delta S_2^{KW} \leq \Delta S_2^{kink}$.
{ Comparison with Fig.~\ref{fig:Asymmetry_in_preserving_cases.} also shows that, as expected, as soon as kink production is increased, kink-entanglement entropy asymmetry is enhanced. In addition, we see that the KW kink asymmetry $\Delta S_2^{KW}$ is in better agreement with $\Delta S_2^{kink}$. This feature can be understood as due to the contribution from possible kinks precisely at the interface playing a less dominant effect.}

\section{Conclusion and Outlook}

In summary, we studied entanglement associated with meson dynamics via confined string evolution and effects of string breaking; we quantified these phenomena through a novel application of entanglement asymmetry in a QCD-analogue model system, a transverse-field Ising chain with three-spin interaction and longitudinal field. We performed the R\'enyi entropy calculation directly in the two-kink subspace. For kink-preserving dynamics, the absence of a longitudinal field gives rise to integrability that sets an upper bound on R\'enyi entropy. We find that the dynamics of entanglement production generically involves two stages: when the string is contracting, entropy is reduced, followed by an increase when the minimum size is reached. Turning on the confining field can break the integrability bound with weak confining field. However, the R\'enyi entropy is suppressed upon further increasing the strength of the confining field. This calculation reveals the internal dynamics of a simulated bound-state system resembling the meson in QCD through the lens of entanglement entropy.

Another significant aspect of our work is the study of the interplay of entanglement and kink production in the context of entanglement asymmetry. To do so, we introduced the kink entanglement asymmetry and the  Kramers-Wannier entanglement asymmetry to address the nature of kink number in our spin chain. To study these numerically, we devised a new calculation of R\'enyi asymmetry using MPS methods and demonstrated its application in the context of kink entanglement asymmetry. 

We comment on a few future directions that are worthy of consideration. A few recent papers \cite{rath2023entanglement, vittorio2022symmetry, neven2021symmetry} explored and proposed measuring entanglement asymmetry in quantum hardware simulations, and it will be interesting to simulate new types of entanglement asymmetry on NISQ devices. In addition, there is a need to develop a more computationally efficient approach to obtain R\'enyi asymmetry (see e.g. \cite{feldman2022entanglement}). Addressing this question might allow us to access later time dynamics for R\'enyi asymmetry, and to see if dynamical purification \cite{vittorio2022symmetry} can be observed in settings similar to our setup. Finally, lattice gauge theories with site and link variables provide natural playgrounds to explore the concept of kink entanglement asymmetry proposed in this work.

{\bf Acknowledgements}
IK would like to thank German Sierra and Sara Murciano for helpful remarks. B.J.J.K.~thanks Joseph Vovrosh for useful discussions about his work. DMK would like to thank Alessandro Roggero for fruitful comments. This document was prepared using the resources of the Fermi National Accelerator Laboratory (Fermilab), a U.S. Department of Energy (DOE), Office of Science, HEP User Facility. Fermilab is managed by Fermi Research Alliance, LLC (FRA), acting under Contract No. DE-AC02-07CH11359. The work of IK was supported in part by the NSF grant DMR-1918207. B.J.J.K., D.M.K., T.J.H., and G.N.P.~were partially supported by the DOE HEP QuantISED grant KA2401032. The work of T.J.H.~at Argonne National Laboratory was supported by the U.S.~Department of Energy under contract DE-AC02-06CH11357. D.M.K. acknowledges partial support by the U.S. Department of Energy, Office of Science, National Quantum Information Science Research Centers, Superconducting Quantum Materials and Systems Center (SQMS) under the contract No. DE-AC02-07CH11359. The tensor network numerics were implemented using the iTensor package~\cite{itensor, itensor-r0.3}.

\bibliography{ref.bib}

\onecolumngrid

\appendix

\section{Benchmarking exact diagonalization, two-kink dynamics and tensor network simulations}

In this appendix, we present numerical data benchmarking. (1) We study the time evolution dynamics of the spin, initial domain wall, and R\'enyi entropy for both the exact diagonalization and the two-kink Hamiltonian dynamics, both at the special point $J = - g$ (where they should match) and $J \neq - g$ (where they generally differ). (2) We compare the time evolution dynamics for the initial domain wall and R\'enyi entropy for tensor network simulation and two-kink dynamics for larger system size.

\begin{figure}[h]
\centering
\includegraphics[width=1.\textwidth]{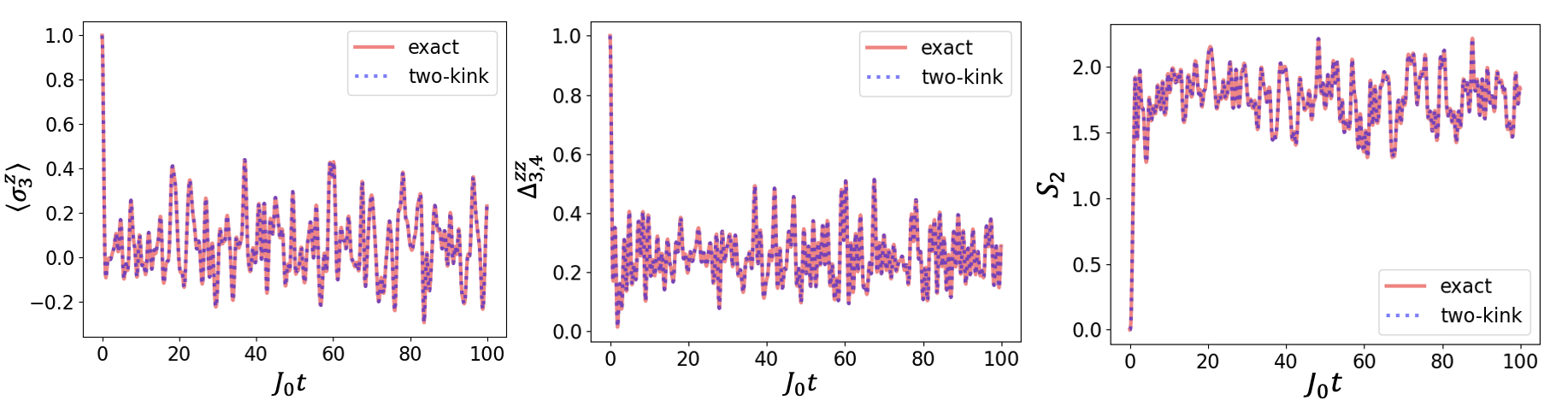}
    \caption{From left to right: The spin expectation value $\langle \sigma^z_3 \rangle$, the domain wall expectation value $\Delta_{3,4}^{zz}$ and the half chain R\'enyi entropy $S_2$ at parameter values $g = 0.7 = - J, h = 0.1$ for both exact and two-kink dynamics. At this special point where $g = -J$, the time evolution dynamics for the exact diagonalization can be described exactly by two-kink subspace dynamics.}
\label{fig:exact_tk_match}
\end{figure}

\begin{figure}[h]
\centering
\includegraphics[width=1.\textwidth]{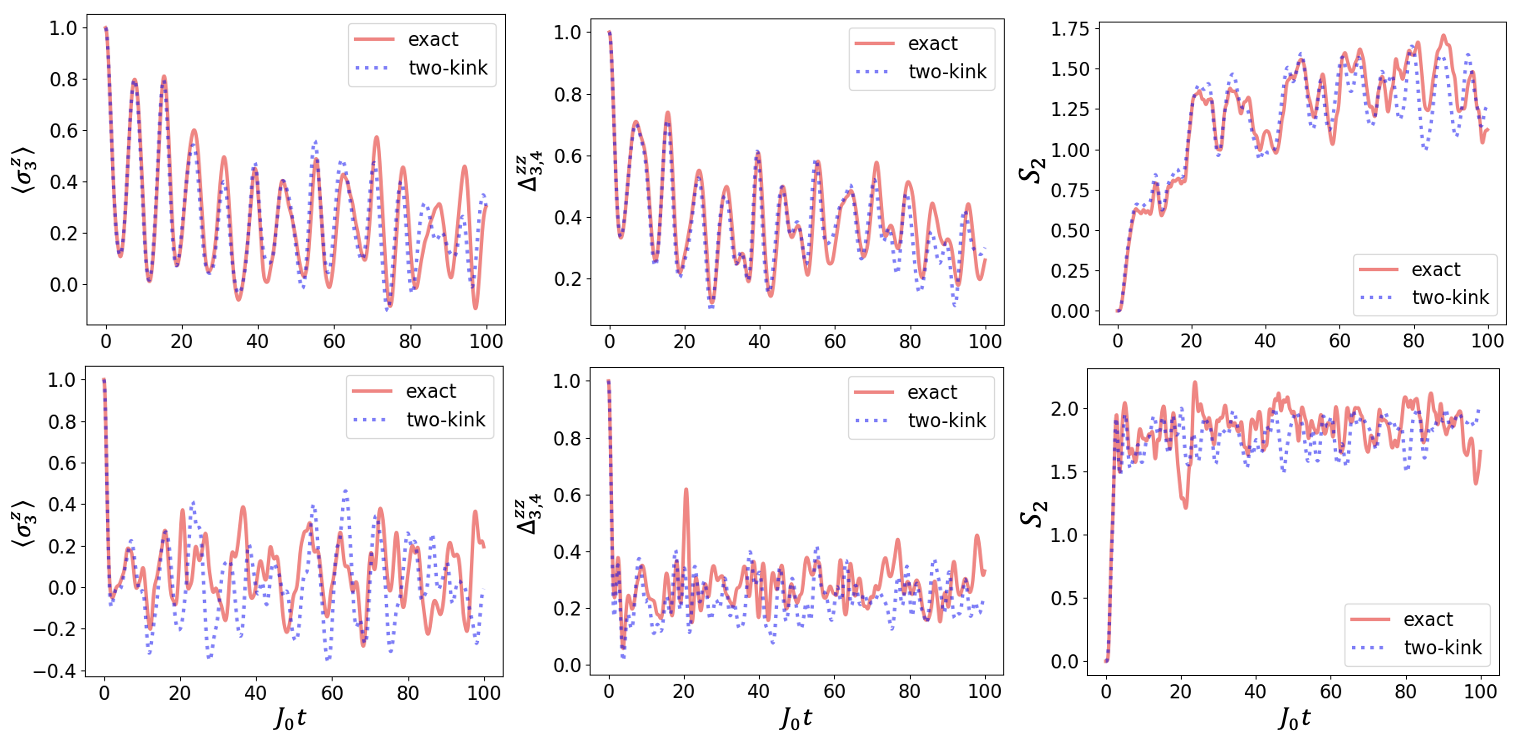}
    \caption{From left to right: The spin expectation value $\langle \sigma_3^z \rangle$, the domain wall expectation value $\Delta_{3,4}^{zz}$ and the half chain R\'enyi entropy $S_2$. From top to bottom: The set of physical quantities at the top panel was simulated at $g = 0.3, h = 0.4$ and $J = -0.05$ while the bottom panel was at $g = 0.6, h = 0.1, J = -0.1$. Note that the top panel corresponds to a more confined dynamics, where the deviation of the time evolution and exact diagonalization occurs only at late time, while for parameter with less confinement (smaller $h$), the deviation is significant even at early times.}
\label{fig:exact_tk_deviate}
\end{figure}

For the first benchmarking between exact diagonalization and two-kink dynamics (to establish the reliability of two-kink dynamics in kink-conserving cases), we initialize the quantum state of a $L = 10$ quantum Ising spin chain to $| \psi_0 \rangle = | \uparrow \uparrow \uparrow \uparrow \downarrow \downarrow \uparrow \uparrow \uparrow \uparrow \rangle$, and time evolve the state using (a) exact diagonalization of the Hamiltonian Eq. 1 in the main text and (b) two-kink Hamiltonian  Eq. 4 in the main text. For Fig.~\ref{fig:exact_tk_match} and \ref{fig:exact_tk_deviate}, the spin expectation values $\langle \sigma^z_i (t) \rangle$ are tracked at the underlined spin $| \uparrow \uparrow \uparrow \underline{\uparrow} \downarrow \downarrow \uparrow \uparrow \uparrow \uparrow \rangle$ in the initial state $| \psi_0 \rangle$, and the domain wall expectation values $\Delta^{zz}_{i,i+1} (t)$ are tracked at $| \uparrow \uparrow \uparrow \underline{\uparrow \downarrow} \downarrow \uparrow \uparrow \uparrow \uparrow \rangle$, and the R\'enyi entropy is cut at half chain bipartition respectively. 

In Fig.~\ref{fig:exact_tk_match}, we see that the time evolution of the various different physical quantities of interest match between exact diagonalization and two-kink subspace evolution. As demanded by the symmetry of the Hamiltonian when $g = -J$, the domain wall number at this special line is exactly conserved.

We then turn to the string breaking (more generic) cases, where kink number is not conserved. In Fig.~\ref{fig:exact_tk_deviate}, we observe that while we generally expect the dynamics of the exact Hamiltonian to differ from that from the two-kink subspace projection at $g \neq - J$, the macroscopic expectation values for spin $\langle \sigma^z_i \rangle$ and domain wall $\Delta^z_{i,i+1}$ overlaps significantly for early times and only deviate at later time for strong confining field $h$. In contrast, with weak confining field $h$ the dynamics deviates significantly at early times. This establishes the general physics and intuition that with strong confinement $h$, two-kink approximation is reasonable for quantum quench problem with initial state $| \uparrow \dots \uparrow \downarrow \dots \downarrow \uparrow \dots \uparrow \rangle$.

\begin{figure}[h]
\centering
\includegraphics[width=0.9\textwidth]{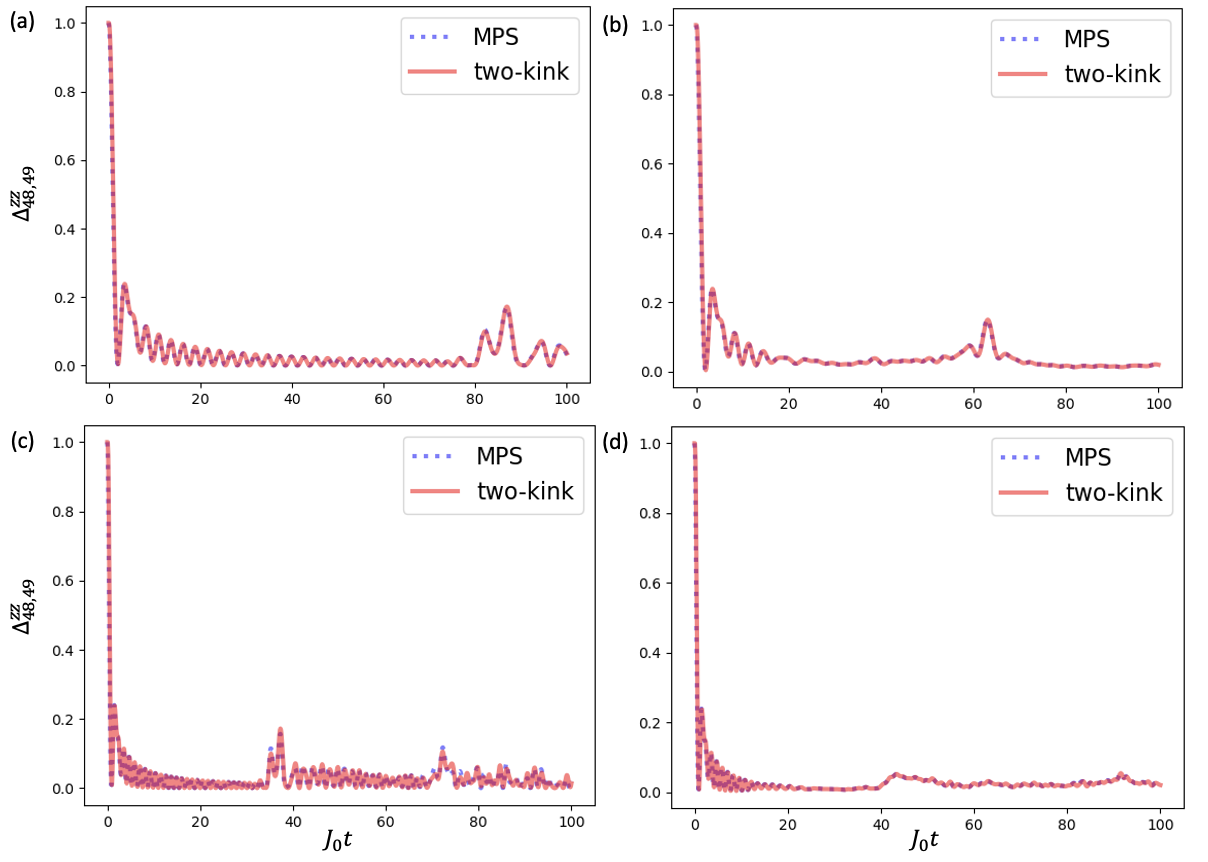}
    \caption{The set of parameter values used for the time evolution of the domain wall $\Delta_{48,49}^{zz}$ at position 48 and 49 following a quantum quench are (a) $g = 0.3, J = -0.3, h = 0$, (b) $g = 0.3, J = -0.3, h = 0.05$, (c) $g = 0.7, J = -0.7, h = 0.0$, (d) $g = 0.7, J = -0.7, h = 0.05$.}
\label{fig:kink-preserving-domain}
\end{figure}

After verifying the validity of the two-kink dynamics for kink-conserving dynamics, we then turn to the benchmarking of the tensor network simulations with the two-kink dynamics. In the main article, we provided the benchmarking of MPS with two-kink dynamics for the time evolution of the R\'enyi entropy $S_2$ for $g = -J = 0.7$ in Fig. 1 in the main text. Here, we provide additional numerical simulations of the domain wall profile located at the initial domain wall positions at $48^{\textrm{th}}$ and $49^{\textrm{th}}$ sites for a spin chain of length $L = 100$ and initial state with 4 middle flipped spin $| \uparrow_1 \dots \uparrow_{48} \downarrow_{49} \downarrow_{50} \downarrow_{51} \downarrow_{52} \uparrow_{53} \dots \uparrow_{100} \rangle$, as shown in Fig.~\ref{fig:kink-preserving-domain}. This provides an additional check (besides the R\'enyi entropy dynamics) that the MPS time evolution from TEBD is approximated well by two-kink time evolution.

\section{Construction of the reduced density matrix in the Two-Kink Subspace}

In this section, we outline the overall approach we used to construct the reduced density matrix in the two-kink subspace in order to calculate the 2$^{\textrm{nd}}$ order R\'enyi entropy for dynamics within the two-kink subspace. We represent all possible two-kink states in a spin chain of length $L$, $| \uparrow_1 \dots \uparrow_{j_L - 1} \downarrow_{j_L} \dots \downarrow_{j_R} \uparrow_{j_R + 1 } \dots \uparrow_L \rangle$, with the two number representation $| j_L, j_R \rangle$, where $j_L$ is the position of the left domain wall and $j_R$ is the position of the right  domain wall. In the two-kink Hilbert space, for sites $i = 1, \dots, L$, the two numbers can take the following values $1 < j_L \leq j_R < L$.

With this two-number representation, a quantum spin chain of size $L$ will have a two-kink subspace of dimension $(L-1)(L-2)/2$. To proceed with the R\'enyi entropy calculation, given a bipartition bond $l_B$ separating the spin chain into left-right bipartition with sites $1, \dots, l_B$ to the left side of the chain and sites $l_B + 1, \dots, L$ to the right, we will proceed to perform bipartition in the two-kink states
\begin{eqnarray}
    | \psi \rangle &=& \sum_{1 < j_L \leq j_R \leq l_B < L} \alpha_{j_L, j_R} | \text{one/two-kink} \rangle_L | \text{no-kink} \rangle_R + \sum_{1 < j_L \leq l_B < j_R < L} \alpha_{j_L, j_R} | \text{one-kink} \rangle_L | \text{one-kink} \rangle_R \nonumber \\ 
    &+& \sum_{1 \leq l_B < j_L \leq j_R < L} \alpha_{j_L, j_R} | \text{no-kink} \rangle_L | \text{one/two-kink} \rangle_R , 
    \label{eq: two-kinks decomposition}
\end{eqnarray}
where we decompose the sum into three types of terms classified by the location of the bipartition cut relative to the domain wall positions. Here, the states with label $| \text{one-kink} \rangle_L | \text{no-kink} \rangle_R$ and $| \text{no-kink} \rangle | \text{one-kink} \rangle_R$ are cases where the bipartition cut $l_B$ coincides with one of the left/right domain walls. 

Constructing the full pure state density matrix $| \psi \rangle \langle \psi |$ and taking the partial trace over the right partition, we have
\begin{eqnarray}
    \rho_L &=& \text{Tr}_R (| \psi \rangle \langle \psi |) = \sum_{1 < j_L \leq j_R \leq l_B } \sum_{1 < j_L' \leq j_R' \leq l_B} \alpha_{j_L, j_R} \alpha_{j_L', j_R'}^* | \text{one/two-kink} \rangle \langle \text{one/two-kink} | \nonumber \\
    &+& \sum_{1 < j_L \leq l_B} \sum_{1 < j_L' \leq l_B} \left( \sum_{l_B < j_R < L} \alpha_{j_L, j_R} \alpha_{j_L', j_R}^* \right) | \text{one-kink} \rangle \langle \text{one-kink} | \nonumber \\
    &+& \left( \sum_{l_B < j_L \leq j_R < L} \alpha_{j_L, j_R} \alpha_{j_L, j_R}^* \right) | \text{no-kink} \rangle \langle \text{no-kink} | \nonumber \\ 
    &+& \sum_{1 < j_L \leq l_B} \left( \sum_{l_B < j_R < L} \alpha_{j_L, j_R} \alpha_{j_L'=l_B+1, j_R}^* \right) | \text{one-kink} \rangle \langle \text{no-kink} | \nonumber \\
    &+& \sum_{1 < j_L' \leq l_B} \left( \sum_{l_B < j_R < L} \alpha_{j_L', j_R}^* \alpha_{j_L=l_B+1, j_R} \right) | \text{no-kink} \rangle \langle \text{one-kink} | 
    \label{eq: reduced density matrix result}
\end{eqnarray}
We shall elaborate on each term below.

\subsection*{The first term}

The first term in Eq.~(\ref{eq: reduced density matrix result}) is obtained from the multiplying the first terms of Eq.~(\ref{eq: two-kinks decomposition}) in both $| \psi \rangle$ and $\langle \psi |$, followed by partial right trace over all spin-up state $| \text{no-kink} \rangle_R$. Hence, the coefficients in the first term in Eq.~(\ref{eq: reduced density matrix result}) can be understood to be constructed without performing any sum from partial tracing.

\subsection*{The second term}

The second term in Eq.~(\ref{eq: reduced density matrix result}) is obtained by multiplying the second terms of Eq.~(\ref{eq: two-kinks decomposition}) in both $| \psi \rangle$ and $\langle \psi |$, followed by partial right trace over $| \text{one-kink} \rangle_R$. In this case, the coefficients in the second term in Eq.~(\ref{eq: reduced density matrix result}) is constructed from summing over $l_B < j_R = j_R' < L$ where the right partition states $| \text{one-kink} \rangle$ and $\langle \text{one-kink} |$ must agree in the partial trace process. 

We also note here that terms labelled as 'the second term' is actually a subset of 'the first term' and not a new set of distinct terms (which we label separately only because they had different origin), so the coefficients here should add to the coefficients of the first term when the states are the same.

\subsection*{The third term}

The third term in Eq.~(\ref{eq: reduced density matrix result}) is actually a single state, and its coefficient is constructed from a sum over the modulo square $| \alpha_{j_L, j_R} |^2$ for $l_B < j_L \leq j_R < L$. Here $j_L$ and $j_R$ both lie on the right partition basis states, for which the indices must match (hence mod square) when one performs the partial trace.

\subsection*{The fourth and fifth terms}

The fourth and the fifth term in Eq.~(\ref{eq: reduced density matrix result}) are the complex conjugate of each other, so it suffices to explain one of them. The fourth term comes from the cross multiplication of the second term $| \text{one-kink} \rangle | \text{one-kink} \rangle$ and the third term of type $\langle \text{no-kink} | \langle \text{one-kink} |$ in Eq.~(\ref{eq: two-kinks decomposition}). 

Note that the third term of the form  $\langle \text{no-kink} | \langle \text{two-kink} |$ cannot contribute since the right partition partial cannot match those from $| \text{one-kink} \rangle | \text{one-kink} \rangle$. This explains the setting of $j_L' = l_B + 1$ in the coefficient of the fourth term (likewise the corresponding term in the fifth term). We also understand that there is a single sum matching the indices $j_R = j_R'$ over the range $l_B < j_R < L$ when one construct the coefficients of the fourth and the fifth term.

\subsection*{Matrix representation of the reduced density matrix}

Collecting these terms, it helps to visualize these various terms on a matrix representation. We label row sectors according to $| \text{two-kink} \rangle$, $| \text{one-kink} \rangle$ and $| \text{no-kink} \rangle$ and likewise for the column sectors. The matrix representation is as below.
\begin{eqnarray*}
    \rho_L = 
    \left( \begin{array}{@{}c|c@{}|c@{}}
    \begin{matrix}
        \text{First term} \\
        | \text{two-kink} \rangle \langle \text{two-kink} |  
    \end{matrix} 
    & 
    \begin{matrix}
        \text{First term} \\
        | \text{two-kink} \rangle \langle \text{one-kink} |  
    \end{matrix} 
    & 0  \\
    \cmidrule[0.4pt]{1-3}
    \begin{matrix}
        \text{First term} \\
        | \text{one-kink} \rangle \langle \text{two-kink} |  
    \end{matrix} 
    & 
    \begin{matrix}
        \text{First + Second term} \\
        | \text{one-kink} \rangle \langle \text{one-kink} |  
    \end{matrix} 
    & 
    \begin{matrix}
        \text{Fourth term} \\
        | \text{one-kink} \rangle \langle \text{no-kink} |  
    \end{matrix} 
    \\
    \cmidrule[0.4pt]{1-3}
    0  & 
    \begin{matrix}
        \text{Fifth term} \\
        | \text{no-kink} \rangle \langle \text{one-kink} |  
    \end{matrix}
    & 
    \begin{matrix}
        \text{Third term} \\
        | \text{no-kink} \rangle \langle \text{no-kink} |  
    \end{matrix}
    \\
    \end{array} \right)
\end{eqnarray*}
Here, the third row is a single row and the third column is a single column. The reduced density matrix can then be used to calculate the R\'enyi entropy for the two-kink dynamics.

\section{R\'enyi Entropy Evolution and Collisions: Early time and Entropy bound violation with large system size} 

\begin{figure}[h]
\centering
\includegraphics[width=0.7\textwidth]{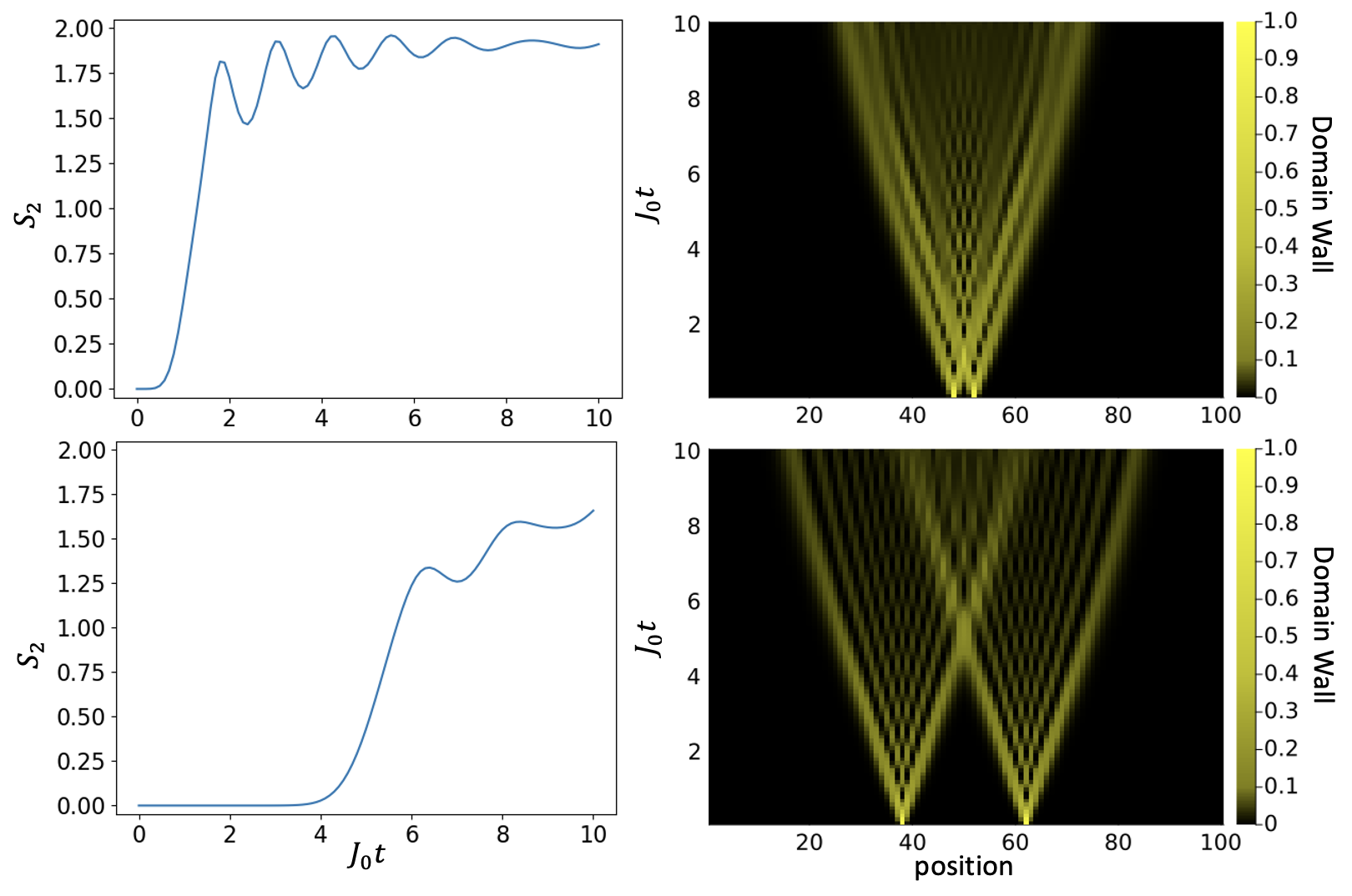}
    \caption{The half-chain R\'enyi entropy $S_2$ (left) and the domain wall profile across link positions (right) for the parameter value set $g= -J= 0.7, h=0.1$. Here the top diagrams are results from initial state with domain wall separation $l = 4$ (4 down spins in the middle) while the bottom diagrams are results with initial domain wall separation $l = 24$. The increase in $S_2$ in the early time roughly coincides with the time scale of collision for kink-conserving dynamics. The spin chain is of length $L = 100$.}
\label{fig:early_time}
\end{figure}

In this appendix, we cover two points. (1) We outline several qualitative features of the R\'enyi entropy evolution that highlight several interesting features that are not discussed in the main article. (2) In addition, we show that the entropy bound violation with confining field is independent of the finite system size effect. We will mainly concern ourselves with the kink-preserving dynamics in this appendix.

For generic model parameters, as the initial product state domain wall starts evolving, entanglement immediately appears, dominated by the dynamics of kink-anti kink pair creation and motion. In contrast, for kink-number preserving dynamics, the half-chain entanglement can only begin emerging when the kinks collide. To see this, we compute the R\'enyi entropy of half of the chain. In Fig.~\ref{fig:early_time}, we show how a rapid increase in the half-chain R\'enyi entropy $S_2$ for small domain wall and large domain wall separation is  associated to the time of the domain wall spreading and collision. Here, a small initial domain wall separation ($l=4$) has a collision around $t\approx 1$ while that of the large separation $l = 24$ has collision around $t \approx 5$. 

\begin{figure}[h]
\centering
\includegraphics[width=0.85\textwidth]{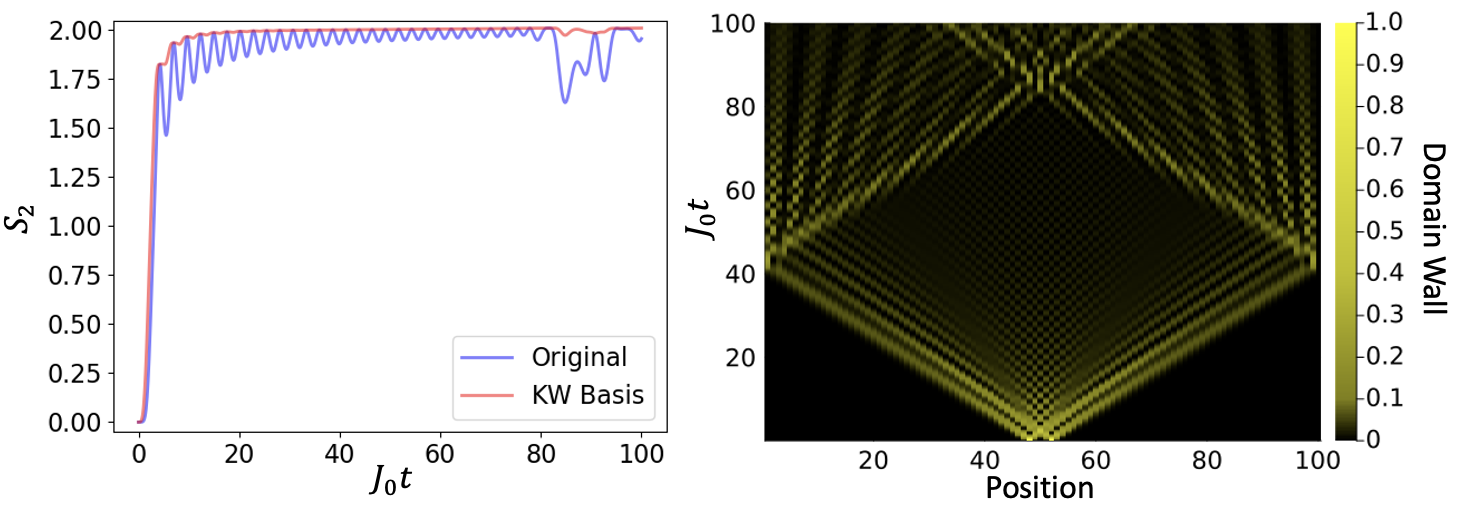}
    \caption{The half-chain R\'enyi entropy $S_2$ (left) and domain wall heat (right) for $g = -J = 0.3$, $h = 0$, $L = 100$, and initial domain wall separation size $l = 4$. The dip around $t \approx 85$ corresponds to kink collisions. The color scale has been magnified near $\Delta^{zz}_{i,i+1} = 0$ as the kink density becomes diluted upon spreading.}
\label{fig:collision 1}
\end{figure}

We provide additional details for the R\'enyi entropy dips found in Fig. 5 in the main text. Here, the dip around $t \approx 85$ is directly associated with kink collision after bouncing off the open boundary chain. We provide the associated domain wall/kink heatmap in Fig.~\ref{fig:collision 1}.

We now turn to demonstrate that the violation of the $S_2 = 2$ bound is independent of the system size, along with other qualitative features of the entropy evolution. In Fig.~\ref{fig:big_size}, we vary various parameters within the kink-conserving dynamics: (1) the kinetic energy $(g - J)$, (2) the confining potential field $h$, and (3) the system size $L$. 

\begin{figure}[h]
\centering
\includegraphics[width=0.9\textwidth]{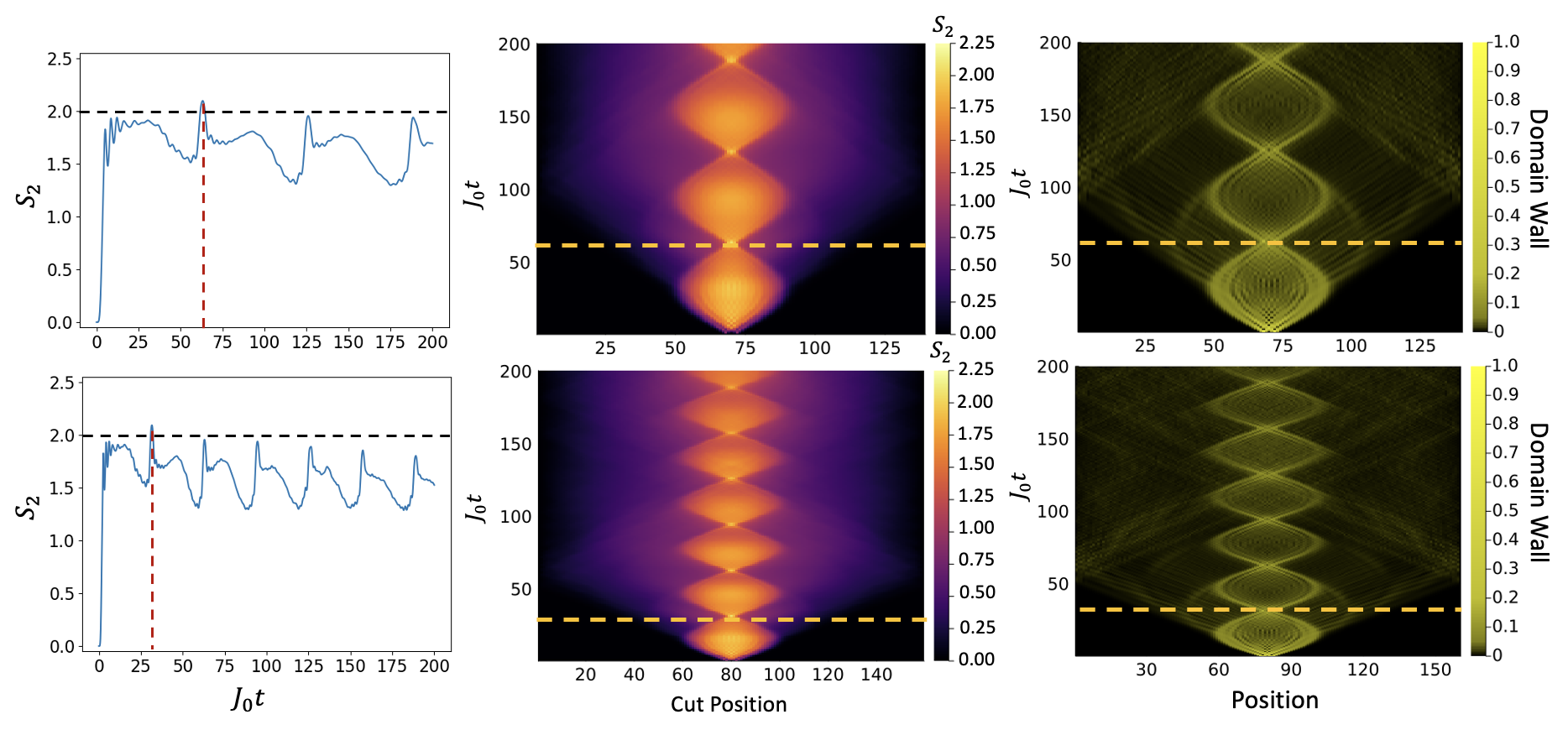}
    \caption{The half-chain R\'enyi entropy $S_2$ (left), the R\'enyi entropy map with different bipartite cut position (middle), and the domain wall profile across link positions (right) for $g = -J = 0.25, h = 0.05, L=140$ (top) and $g = -J = 0.5, h = 0.1, L=160$ (bottom).}
\label{fig:big_size}
\end{figure}

In Fig.~\ref{fig:big_size}, we see the $S_2$ entropy bound is violated upon the second collision of the kinks. The first collision happens at early time and is discussed in detail earlier in this appendix. Here, we verify that the total kink density near the ends of the chain is negligible when the second collision occurs upon the acceleration of the kink dynamics back to the middle of the chain owing to the confining potential.  

Another feature of the kink-conserving dynamics is the dip in the R\'enyi entropy before the periodic collisions of the kinks after expanding to maximal extent set by the strength of the confining potential. This can be observed in the half-chain R\'enyi entropy, as well as in the R\'enyi entropy heatmap diagram, where the color code shows a lower R\'enyi entropy dip right before the periodic collisions of the kinks.

\section{MPS Implementation and Computational Complexity}

Many $1D$ systems can be analyzed efficiently using matrix product states (MPS). Previous work has explored entanglement asymmetry in the context of exact diagonalization \cite{ares2023entanglement}, where large system sizes can only be explored for non-interacting models, and iTEBD time evolution for interacting models with integrability \cite{bertini2023dynamics}. Another recent work \cite{capizzi2023universal} has explored entanglement asymmetry in matrix product states in the context of the ground state of a symmetric Hamiltonian undergoing spontaneous symmetry breaking. Here we implement the computation of the $\Delta S_2$ entanglement asymmetry given an MPS for generic models that includes non-integrability, an interacting Hamiltonian, and a more general setting that includes non-local entanglement asymmetry.

We will now outline the general algorithm for the straightforward computation of the bipartite entanglement asymmetry for MPS.
\begin{enumerate}
    \item 
    Perform the time evolution of the MPS from $| \psi(t-\Delta t) \rangle$ to $| \psi (t) \rangle$ with the choice of tMPS, tDMRG or TEBD.

    \item 
    Choose the orthogonality center of the MPS across which the bipartite 2nd R\'enyi entropy $S_2 (\rho_A)$ is calculated (mixed canonical MPS).

    \item 
    Contract the indices efficiently (with increasing bond dimensions) from one end of the MPS to the orthogonal center to construct the reduced density matrix $\rho_A$ (Fig.~\ref{fig:MPS algorithms.}(a)).

    \item 
    With $\rho_A$, split the numerical integration of Eq.~15 in the main text into $k+1$ steps. For each discrete $\lambda$ perform the MPO application onto the $\rho_A$ [Fig.~\ref{fig:MPS algorithms.}(b) or (c)]. Perform the full trace after evaluating $e^{i \lambda Q} \rho_A e^{- i \lambda Q} \rho_A$. The MPO application and tracing is done $k+1$ times to evaluate $S_2(\rho_{A, Q})$ and subsequently $\Delta S_2$.
\end{enumerate}

In this work, we calculate two different types of entanglement asymmetry: onsite entanglement asymmetry (such as the magnetization entanglement asymmetry previous considered in \cite{ares2023entanglement} with $Q_A = \sum_{i=2}^l \sigma_i^z$) and link-type entanglement asymmetry (in our case, the kink entanglement asymmetry with $Q_A = \sum_{i=1}^{l-1} \sigma_i^z \sigma_{i+1}^z$, shown in Fig.~\ref{fig:MPS algorithms.}(c)).

\begin{figure}[h]
\centering
\includegraphics[width=0.7\textwidth]{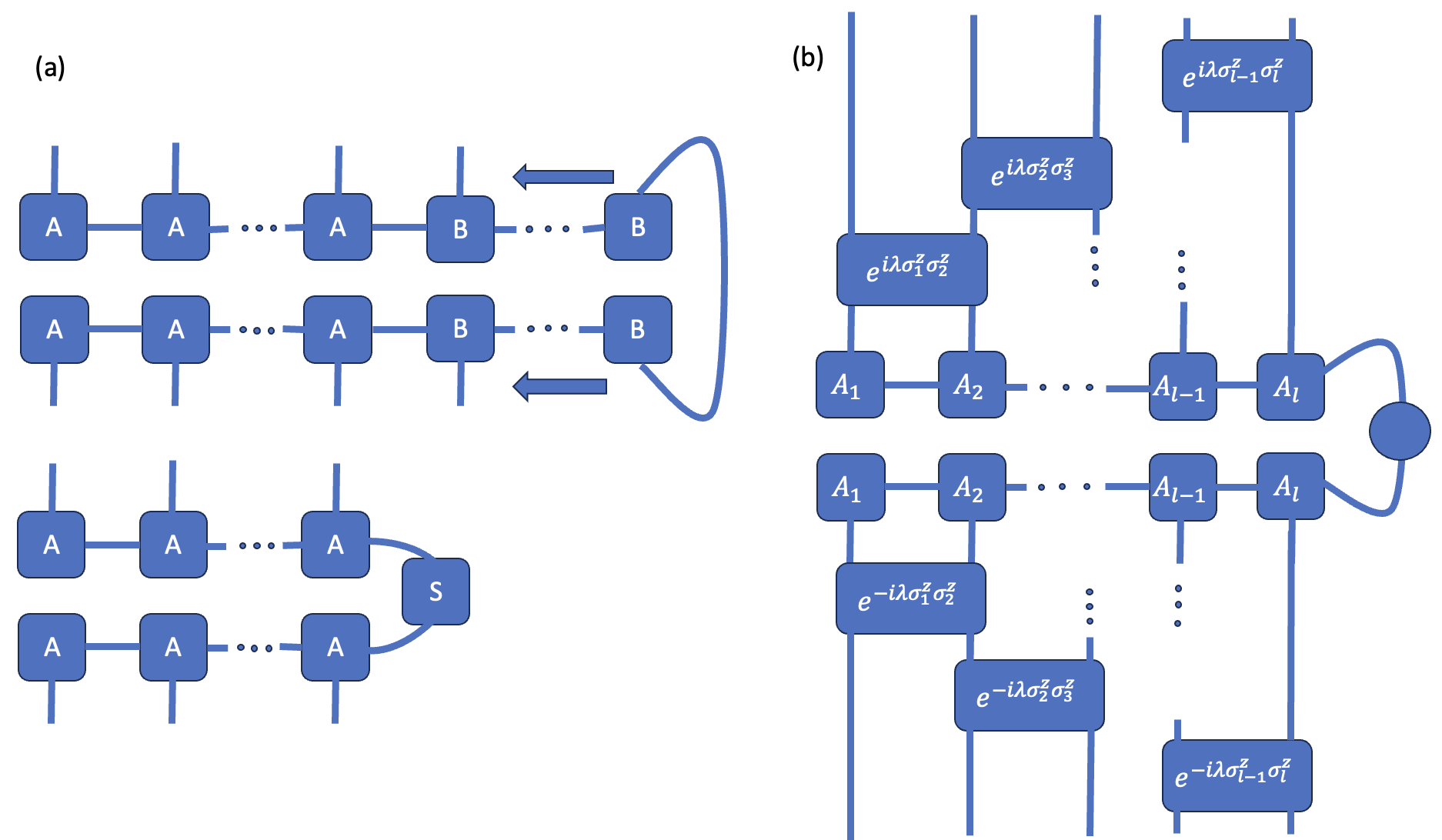}
    \caption{(a) The order of contraction of the legs of the tensor of the MPS wave function in the construction of the reduced density matrix $\rho_A = \text{Tr}_B (| \psi_{MPS} \rangle \langle \psi_{MPS} |)$ is shown in the arrow. (b)The MPO operators $e^{\pm i \lambda Q} = \prod_{i=1}^{l} e^{\pm i \lambda \sigma_i^z}$ (in site-type symmetry resolved entropy), and (c) The MPO operators $e^{\pm i \lambda Q} = \prod_{i=1}^{l-1} e^{\pm i \lambda \sigma_i^z \sigma_{i+1}^z}$ (in link-type symmetry resolved entropy) are applied successively via the DMRG algorithm. If the integral is split into $k+1$ steps, $\lambda$ takes discrete values in the interval $[- 2 \pi, 2 \pi ]$ and this calculation is done for $(k+1)$ different $\lambda$ before the Trapezoid Rule is applied to obtain $\rho_{A, Q}$}
\label{fig:MPS algorithms.}
\end{figure}

\begin{figure}[h]
\centering
\includegraphics[width=0.7\textwidth]{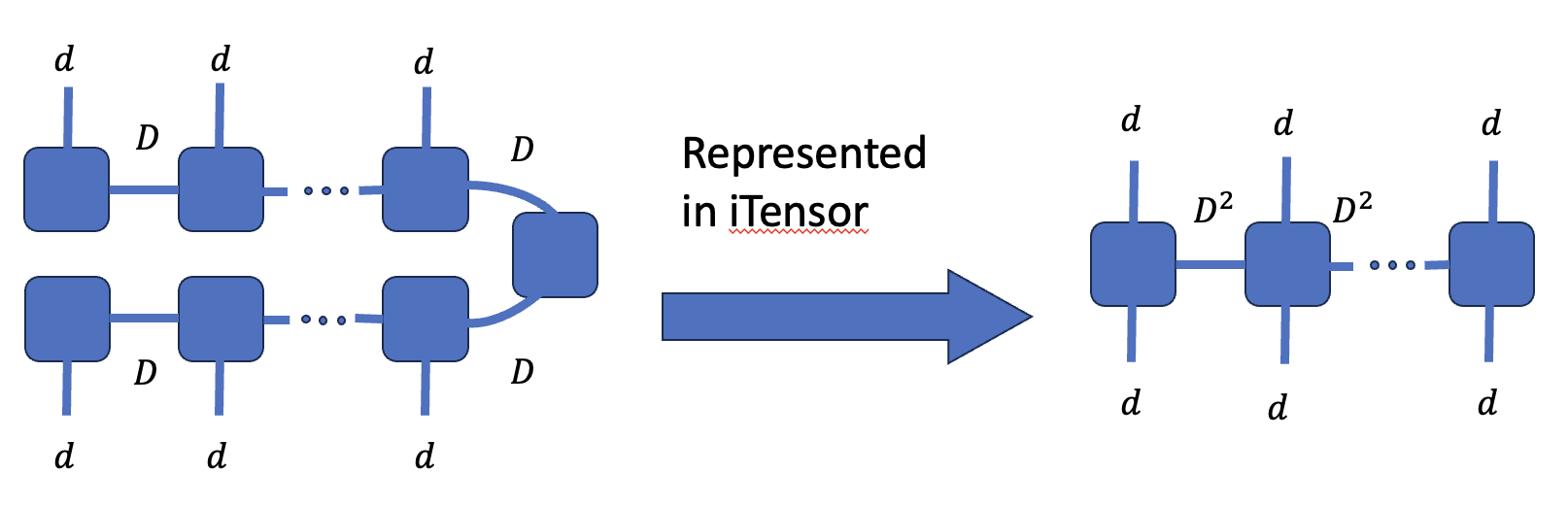}
    \caption{In iTensor, taking an outer product of an MPS $| \psi \rangle$ with itself (with bond dimension $D$) automatically forms an MPO density matrix with bond dimensions $D^2$.}
\label{fig:rho itensor}
\end{figure}

\begin{figure}[h]
\centering
\includegraphics[width=0.9\textwidth]{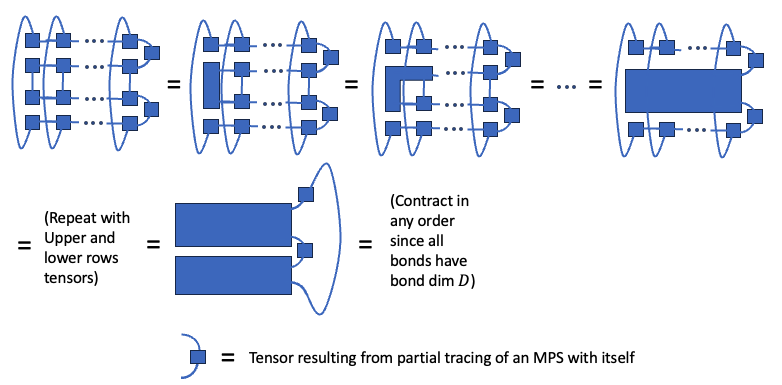}
    \caption{Minimal contraction order for individual tensors to compute $\text{Tr}(e^{i \lambda Q} \rho_A e^{- i \lambda Q} \rho_A)$. Here the top 2 and bottom 2 rows of tensors can come from different reduced density matrix. The strategy for implementation here is to perform contractions of individual tensor component in MPS wave functions without first forming the density matrix MPOs, and the size tensor from partial trace can also be formed partial inner product of MPS wave function with itself.}
\label{fig:Efficient contraction}
\end{figure}

We will comment on the computational complexity of our algorithm above, and possible speed up that could possibly be implemented in a future work. In iTensor, working with density matrices is actually sub-optimal since it creates MPO with bond dimension $D^2$, which we illustrate in Fig.~\ref{fig:rho itensor}. Depending on the details of how iTensor package handles the contraction of two (different) reduced density matrices, this computation can either scale as $\mathcal{O} (L d^3 D^8)$ in the worst case or $\mathcal{O} (L d^3 D^6)$ in the most optimal case. Dealing with reduced density matrices directly is more straightforward in code development with existing methods in iTensor, but severely limits the simulability of R\'enyi entropy evolution to later times when bond dimension is large everywhere on the spin chain. This is the case for string breaking (kink number violating, generic case) situation when $g \neq -J$, as shown in Fig. 6 in the main text when we attempt to calculate R\'enyi asymmetry directly with density matrix contractions and issue with computational cost only allows time evolution for early times ($t = 10$).

Instead, an optimal approach is shown in Fig.~\ref{fig:Efficient contraction}. The algorithm represented pictorially in Fig.~\ref{fig:Efficient contraction} will necessitate contractions without constructing reduced density matrices directly. In this approach, we contract tensors individually and avoid the dealing with the full $D^2$ bond dimensions of each $\rho_A$ MPO link. The algorithm in this case will dramatically improves the complexity to $\mathcal{O} (L d D^3)$, but it comes with a more complicated code development. We will leave the development of a more efficient entanglement asymmetry computation as a future research avenue.

\section{Kramers-Wannier Unitary Transformation and the XY Model}

Given the problem that the kink entanglement asymmetry at the interface cannot be tracked in our current basis, we perform a site-to-link transformation to map link variables (such as the domain wall operator $\sigma_i^z \sigma_{i+1}^z$) to site variable. This has traditionally been associated to the Kramers-Wannier duality on a 2D Ising Model. In the context of the (1+1)D Transverse Field Ising Model, one cannot construct a unitary that represents the Kramers-Wannier duality since it maps the ferromagnetic phase of the Transverse Field Ising model ($g < 1$) with ground state degeneracy and spin-flip symmetry breaking to the paramagnetic phase of the Transverse Field Ising Model ($g > 1$) with non-degenerate ground state possessing spin-flip symmetry. Hence, we wish to clarify from the onset that our unitary defined below is a Kramers-Wannier unitary transformation that is only well-defined on an open chain and serves to perform a basis transformation.

We define a Kramers-Wannier unitary $U_{KW}$ that is distinct from the Kramers-Wannier duality in the sense that: (1) it is defined on open boundary condition, (2) it maps all links to sites except the first site, which also maps to the first site in the 'dual' lattice, and (3) the unitary is not self-dual, i.e., $U_{KW}^2 \neq I$. In this way the dimension of the Hilbert space is preserved, since an open chain with $N$ sites only has $N-1$ links. The unitary is defined and represented in the quantum circuit language as a series of CNOT gates, as shown in Fig.~\ref{fig:kwucircuit}.

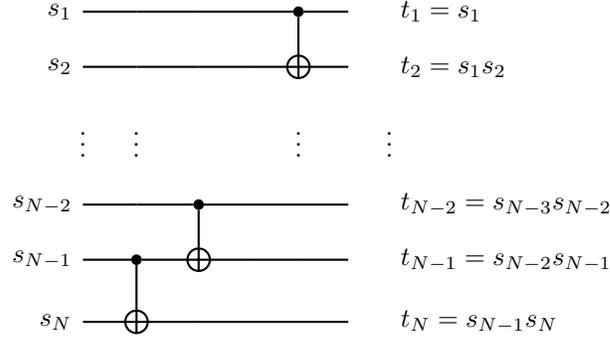
\begin{figure}
\centering
\begin{quantikz}
&\lstick{$s_1$} & \qw & \qw & \qw & \ctrl{1} & \qw &\rstick{$t_1=s_1$} &
\\
&\lstick{$s_2$} & \qw & \qw & \qw & \targ{} & \qw &\rstick{$t_2=s_1 s_2$} & 
\\
& \qvdots  & \qvdots & & & \qvdots &  & \qvdots &
\\
&\lstick{$s_{N-2}$} & \qw & \ctrl{1} & \qw & \qw & \qw &\rstick{$t_{N-2}=s_{N-3}s_{N-2}$} &
\\
&\lstick{$s_{N-1}$} & \ctrl{1} & \targ{} & \qw & \qw & \qw &\rstick{$t_{N-1}=s_{N-2}s_{N-1}$}&
\\
&\lstick{$s_N$} & \targ{} & \qw & \qw & \qw & \qw &\rstick{$t_N=s_{N-1}s_N$}&
&&&&&&&&
\end{quantikz}
\caption{This quantum circuit implements the Kramers-Wannier Unitary on an open Ising spin chain.}
\label{fig:kwucircuit}
\end{figure}

Here, we have defined $| 0 \rangle \equiv | \uparrow \rangle \equiv | s = +1 \rangle$, $| 1 \rangle \equiv | \downarrow \rangle \equiv | s = -1 \rangle$, and one can verify that the action of the CNOT gate on the target qubit/spin matches the value of the product of spin $t_j = s_{j-1} s_j$. In this dual lattice picture, we can perform the unitary transformation on the Hamiltonian $H' = U_{KW} H U_{KW}^{\dagger}$ to get, in this dual picture,
\begin{equation}
    H' = - \left[ J_0\sum^{L}_{i=2} \sigma^z_i + g \sum^{L-1}_{i=2} \sigma^x_i \sigma^x_{i+1} + h \sum^{L}_{i=1} \left( \prod_{j=1}^i \sigma^z_j \right) - J \sum_{i=2}^{L-1} \sigma_i^y \sigma_{i+1}^y \right]\ ,
\label{eq:TFIM_KW}
\end{equation}
where the Kramers-Wannier unitary $U_{KW}$ maps $\sigma_i^z \rightarrow \prod_{j=1}^i \sigma_j^z$, $\sigma_i^x \rightarrow \sigma_i^x \sigma_{i+1}^x$, and the coefficients above identity the origin of the terms of the original Hamiltonian in Eq.~1 in the main text. While the form of the Kramers-Wannier mapping for $\sigma_i^z$ and $\sigma_i^x$ has been given in \cite{ostmann2019localization}, to our knowledge the explicit form of the unitary has not been given in earlier works. In this dual picture, the symmetry operator is transformed as $\prod_{i=1}^{l-1} e^{\pm i \lambda \sigma_i^z \sigma_{i+1}^z} \rightarrow \prod_{i=2}^{l} e^{\pm i \lambda \sigma_i^z}$, the total magnetization of the left half of the spin chain except the magnetization on site $1$. 

Interestingly, while the original Hamiltonian had $4$ sectors associated with local boundary spins $\sigma_1^z$ and $\sigma_L^z$, our transformed Hamiltonian now commutes with $\sigma^z_1$ as well as with the parity operator $\prod_{i=1}^{L} \sigma^z_i$ which is non-local. We also note that the original Hamiltonian preserves the parity of the kink number, even when $g\neq -J$. This form also makes explicit the conservation of kink parity, where kink is now represented by the number of down spins.

When $h\! =\! 0$ in the Hamiltonian Eq.~(\ref{eq:TFIM_KW}), this Hamiltonian is the XY model. This can readily be cast into the free fermion picture using the Jordan Wigner Transformation, with the resulting Hamiltonian (ignoring constant terms)
\begin{equation}
    H = - \left[2 J_0\sum^{L}_{j=2} c_j^{\dagger} c_j + (g+J) \sum^{L-1}_{j=2} (c_j^{\dagger} c_{j+1}^{\dagger} + c_{j+1} c_j ) + (g-J) \sum_{j=2}^{L-1} (c_{j}^{\dagger} c_{j+1} + c_{j+1}^{\dagger} c_j) + h \sum_{j=1}^L \left( \prod_{i=1}^j \left( 2 c_i^{\dagger} c_i - 1 \right) \right) \right]\ .
\label{eq:free_fermion}
\end{equation}
In this form, the fermions represent the domain walls, and we can read off that the hopping strength of the domain wall is $(g-J)$ while the kink number violation term comes with the strength $(g+J)$.

\section{Upper bound on R\'enyi Entropy for $h=0$ and kink-number preservation.}

In this appendix, we prove that starting with a domain wall,  and evolving with our Hamiltonian with $J=-g$ and $h=0$ the second order half chain R\'enyi entropy $S_2 = - \log_2 ( \text{Tr}(\rho_A^2) )$ is bound by $2$ at all times. Concretely, we consider an initial state of the form:
\begin{eqnarray}
    |\psi (0)\rangle =|\uparrow
   \text{...}\uparrow \downarrow
   _x\text{...}\downarrow
   _{y-1}\uparrow _y\text{...}\uparrow
   \rangle
\end{eqnarray}
To compute the evolution, we first apply the Kramers-Wannier $U_{KW}$, mapping the state to:
\begin{eqnarray}
    U_{\text{KW}}|\psi (0)\rangle
   =|1\text{...}(-1)_x111(-1)_y11\text{
   ...}1\rangle .
\end{eqnarray}
Via a Jordan Wigner transformation this state will become the two particle state
\begin{eqnarray}
    U_{\text{KW}}|\psi (0)\rangle
   =c_x^{\dagger }c_y^{\dagger
   }|\text{vac}\rangle. \label{eq: JordanWigner initial state}
\end{eqnarray}
As explained above, the evolution of the state in the Jordan-Wigner picture is governed by the quadratic fermion Hamiltonian \eqref{eq:free_fermion}. In particular, when $g=-J$, the Hamiltonian does not involve pair creation terms. Under such evolution, the creation operators in \eqref{eq: JordanWigner initial state} transform as:
\begin{eqnarray}
    c_{\alpha }^{\dagger }\longrightarrow 
   \Sigma _ju_{\text{$\alpha
   $j}}c_j^{\dagger }
\end{eqnarray}
with a unitary $L\times L$ matrix $u$.
Therefore the evolved state is of the form:
\begin{eqnarray}
    U_{\text{KW}}|\psi (t)\rangle =\Sigma
   _{i,j} u_{\text{xi}  }u_{\text{yj}}c_i^{\dagger
   }c_j^{\dagger }|\text{vac}\rangle
   =\Sigma _{i<j}\left(u_{\text{xi}
   }u_{\text{yj}}-u_{\text{xj}}u_{\text
   {yi}}\right)c_i^{\dagger
   }c_j^{\dagger }|\text{vac}\rangle
\end{eqnarray}
We now undo the Kramers-Wannier transformation to get:
\begin{eqnarray}
    |\psi (t)\rangle =\Sigma
   _{i<j}\left(u_{\text{xi}
   }u_{\text{yj}}-u_{\text{xj}}u_{\text
   {yi}}\right)|\uparrow
   \text{...}\uparrow \downarrow
   _i\text{...}\downarrow
   _{j-1}\uparrow _j\text{...}\uparrow
   \rangle \label{eq:evolved 2 kink state}
\end{eqnarray}
We now proceed by bounding the Schmidt rank of the state \eqref{eq:evolved 2 kink state}. We expand the summation explicitly as:
\begin{eqnarray}  
    |\psi (t) \rangle &=& \Sigma _{i \leq L_A} \Sigma _{j>L_A} \left(u_{\text{xi}} u_{\text{yj}} -u_{\text{xj}} u_{\text{yi}} \right) |\uparrow \text{...} \uparrow \downarrow_i \text{...} \downarrow_{j-1} \uparrow _j \text{...} \uparrow_N \rangle \nonumber \\ 
    &+& \Sigma_{i>L_A} \Sigma_{j>i} \left( u_{\text{xi}} u_{\text{yj}} -u_{\text{xj}} u_{\text{yi}} \right) | \uparrow \text{...} \uparrow \downarrow_i \text{...} \downarrow_{j-1} \uparrow_j \text{...} \uparrow_N \rangle \nonumber \\ 
    &+& \Sigma _{i<j} \Sigma_{j \leq L_A} \left( u_{\text{xi}} u_{\text{yj}} -u_{\text{xj}} u_{\text{yi}} \right) | \uparrow \text{...} \uparrow \downarrow_i \text{...} \downarrow_{j-1} \uparrow_j \text{...} \uparrow_N \rangle \nonumber \\ 
    &=& \left( \Sigma_{i \leq L_A} u_{\text{xi}} | \uparrow \text{...} \uparrow \downarrow_i \text{...} \downarrow _{L_A} \rangle \right) \otimes \left( \Sigma_{j>L_A} u_{\text{yj}} | \downarrow_{L_{A}+1} \text{...} \downarrow_{j-1} \uparrow_j \text{...} \uparrow_N \rangle \right) \nonumber \\ 
    &-& \left( \Sigma_{i \leq L_A} u_{\text{xj}} | \uparrow \text{...} \uparrow \downarrow_i \text{...} \downarrow_{L_A} \rangle \right) \otimes \left( \Sigma_{j>L_A} u_{\text{yi}} | \downarrow_{L_{A+1}} \text{...} \downarrow_{j-1} \uparrow _j \text{...} \uparrow_N \rangle \right) \nonumber \\ 
    &+& |\uparrow \text{...} \uparrow \uparrow_{L_A} \rangle \otimes \left( \Sigma_{i>L_A} \Sigma_{j>i} \left( u_{\text{xi}} u_{\text{yj}} - u_{\text{xj}} u_{\text{yi}} \right) | \uparrow \text{...} \uparrow \downarrow_i \text{...} \downarrow_{j-1} \uparrow _j \text{...} \uparrow_N \rangle \right) \nonumber \\ 
    &+& \left( \Sigma_{i<j} \Sigma_{j \leq L_A} \left( u_{\text{xi}} u_{\text{yj}} - u_{\text{xj}} u_{\text{yi}} \right) | \uparrow \text{...} \uparrow \downarrow_i \text{...} \downarrow_{j-1} \uparrow_j \text{...} \uparrow_{L_A} \rangle \right) \otimes | \uparrow \text{...} \uparrow \uparrow_N \rangle.
\end{eqnarray}
Observe that in the last line we have explicitly separated $  |\psi (t)\rangle $ into a combination of the form:
\begin{eqnarray}
    |\psi (t) \rangle  =\Sigma _{\alpha
   =1}^4|\phi _{\alpha }\rangle \otimes
   |\tilde{\phi }_{\alpha }\rangle. \label{eq:form 4}
\end{eqnarray}
Note that for the rest of the argument, it doesn't matter if  $\phi_{\alpha}, \tilde{\phi}_{\alpha}$, are normalized or represent orthogonal sets. Indeed, the form \eqref{eq:form 4} immediately implies that the Schmidt rank of the state $| \psi (t) \rangle$ with respect to this partition is at most $\text{Sch}(\psi (t))\leq 4$ (see Problem 2.2 in \cite{mikeandike}). Since the logarithm of the Schmidt number is a bound on entropy (Von Neumann entropy, and as consequence also R\'enyi entropy) we have that:
\begin{eqnarray}
    S_2\leq \log _2 \text{Sch}(\psi(t)) \leq \log_2 4 = 2 .
\end{eqnarray}

\end{document}